\documentclass [11pt,a4paper]{article}
\usepackage{jheppub}

\usepackage{graphicx}
\usepackage{color}
\usepackage{amsfonts,amsthm,amssymb,amsmath}
\usepackage{nicefrac}

\def\beq{\begin{equation}}
\def\eeq{\end{equation}}
\newcommand{\be}{\begin{eqnarray}}
\newcommand{\ee}{\end{eqnarray}}

\renewcommand{\texttt}{{}}

\def\bs{\begin{subequations}}
\def\es{\end{subequations}}

\def\la{\lambda}

\def\cF{\mathcal{F}}

\def\Fc{\mathcal{F}}

\def\Hc{\mathcal{H}}

\def\Lc{\mathcal{L}}
\def\Mc{\mathcal{M}}

\def\Pc{\mathcal{P}}

\def\Rc{\mathcal{R}}

\def\Wc{\mathcal{W}}
\def\Zc{\mathcal{Z}}

\def\B{\Box}

\newcommand{\tia}[1]{}

\newcommand{\bea}{\begin{eqnarray}}
\newcommand{\eea}{\end{eqnarray}}
\newcommand{\beas}{\begin{eqnarray*}}
\newcommand{\eeas}{\end{eqnarray*}}
\newcommand{\bal}{\begin{aligned}}
\newcommand{\eal}{\end{aligned}}

\def\({\left(}
\def\){\right)}

\def\la{\lambda}

\newcommand{\pd}{\partial}

\newcommand{\intdg}[1]{\int d^{#1}x\sqrt{-g}}

\newcommand{\cpd}{\nabla}

\newcommand{\const}{\mathrm{const}}

\title{\boldmath $R^2$ inflation to probe non-perturbative quantum gravity}

\author[a,b]{Alexey S. Koshelev,}
\author[a]{K. Sravan Kumar,}
\author[c,d]{and Alexei A. Starobinsky}
\affiliation[a]{Departamento  de F\'isica and Centro  de  Matem\'atica  e 
Aplica\c c\~oes (CMA-UBI),  Universidade  da  Beira  Interior,  6200  Covilh\~a, 
Portugal}
\affiliation[b]{Theoretische Natuurkunde, Vrije Universiteit Brussel, and The International Solvay Institutes, Pleinlaan 2, B-1050 Brussels, Belgium}
\affiliation[c]{L. D. Landau Institute for Theoretical Physics RAS, Moscow 119334, Russian Federation}
\affiliation[d]{Kazan Federal University, Kazan 420008, Republic of Tatarstan, Russian Federation}
\emailAdd{alexey@ubi.pt}
\emailAdd{sravan@ubi.pt}
\emailAdd{alstar@landau.ac.ru}

\abstract{
It is natural to expect a consistent inflationary model of the very early Universe to be an effective theory of quantum gravity, at least at energies much less than the Planck one. For the moment, $R+R^2$, or shortly $R^2$, inflation is the most successful in accounting for the latest CMB data from the PLANCK satellite and other experiments. Moreover, recently it was shown to be ultra-violet (UV) complete via an embedding into an analytic infinite derivative (AID) non-local gravity. In this paper, we derive a most general theory of gravity that contributes to perturbed linear equations of motion around maximally symmetric space-times. We show that such a theory is quadratic in the Ricci scalar and the Weyl tensor with AID operators along with the Einstein-Hilbert term and possibly a cosmological constant. We explicitly demonstrate that introduction of the Ricci tensor squared term is redundant. Working in this quadratic AID gravity framework without a cosmological term we prove that for a specified class of space homogeneous space-times, a space of solutions to the equations of motion is identical to the space of backgrounds in a local $R^2$ model. We further compute the full second order perturbed action around any background belonging to that class. We proceed by extracting the key inflationary parameters of our model such as a spectral index ($n_s$), a tensor-to-scalar ratio ($r$) and a tensor tilt ($n_t$). It appears that $n_s$ remains the same as in the local $R^2$ inflation in the leading slow-roll approximation, while $r$ and $n_t$ get modified due to modification of the tensor power spectrum. This class of models allows for any value of $r<0.07$ with a modified consistency relation which can be fixed by future observations of primordial $B$-modes of the CMB polarization. This makes the UV complete $R^2$ gravity a natural target for future CMB probes. 
}


\begin{document}

\maketitle

\section{Introduction}

Finding a gravity theory consistent with the concepts of quantum field theory is a long-standing problem. General Relativity (GR) \cite{Wald:1984rg} was known not to be ultra-violet (UV) complete from the very beginning. Hence one is forced to modify GR in order to construct any self-consistent model of quantum gravity. Moreover, generalizing GR one has not give it up altogether, as it is heavily supported by absolutely all measurements in the low energy or infra-red (IR) regime including the recent direct discovery of gravitational waves \cite{Abbott:2016blz}.

One of the most obvious and at the same time very promising generalization to consider is $\frac{M_P^2}2\left(R+R^2/(6M^2)\right)$ Lagrangian instead of just $\frac{M_P^2}2R$ as in GR where as usually $R$ is the Ricci scalar, $M_P^{-2}=8\pi G$, $G$ is the Newtonian constant, $c=1$ and $M$ becomes the mass of what is the propagating scalar in this model, dubbed scalaron. We refer it hereafter as the local $R^2$ model or just $R^2$ model (or Lagrangian, etc.). Besides being the simplest one and having only one free parameter $M$ which value is fixed to $M\approx 1.3\times 10^{-5}M_P$ by the
observed Fourier power spectrum $\Pc_{\Rc}$ of primordial scalar (matter
density) perturbations in the Universe, this generalization has two major advantages. First it was proven in \cite{Stelle:1976gc,Stelle:1977ry} that it is renormalizable, i.e. UV complete in the scalar sector. Second, a dramatically successful model of inflation 
\cite{Starobinsky:1980te,Starobinsky:1981vz,Starobinsky:1983zz} known as ``Starobinsky inflation'' is provided by the $R^2$ Lagrangian. With the latest Cosmic Microwave Background (CMB) measurements by the PLANCK mission\cite{Ade:2015xua,Ade:2015lrj} and the more recent BICEP2/Keck Array experiments\cite{Array:2015xqh}, $R^{2}$ inflation produces an excellent fit for the key inflationary predictions.

The advantage of renormalizability of the local $R^2$ gravity has an unfortunate fate to be spoiled by the non-unitarity as a spin-2 ghost appears in the physical spectrum. This ghost is a manifestation of the Ostrogradski instability \cite{Ostro:1850} due to higher derivatives and it appears as long as the required for the full renormalization $W^2$ term, with $W$ being the Weyl tensor, is included. It is not hopeless to try to remove such a ghost and few ways are known in principle. Ghosts may become unphysical in constrained systems \cite{Woodard:2006nt,DeFelice:2010aj}. 
Also one can try to consider special constructions like Horndeski theories \cite{Deffayet:2013lga} in which higher derivatives in the action still result in a second order equations of motion (EOM). Another way is to promote the Lagrangian to a non-local model such that infinitely many derivatives form some operator which does not create new poles in the propagator and consequently does not generate new physical degrees of freedom. On this way possible operators which we may encounter are: analytic in derivatives like $\exp(\Box)$ with $\Box$ being the covariant  d'Alembertian operator, non-analytic in derivatives like $1/\Box$, having logarithms like $\Box\log(\Box)$, etc.

It was shown already in \cite{Starobinsky:1981zc} 
that a systematic accounting of one-loop corrections from quantum matter fields to the $R^2$ gravity leads to infinite derivative logarithmic functions of the d'Alembertian in the action.
Theories with analytic infinite derivative (AID) operators in the action naturally appear in string theory when the string field theory (SFT) \cite{Witten:1986qs,Arefeva:2001ps} is considered. Also $p$-adic string theory \cite{Vladimirov:1994wi} is an example of a model featuring AID Lagrangians. Both of these stringy models are unitary and UV complete theories. A study of gravity theories having similar AID operators was initiated in \cite{Tomboulis:1997gg}. This led recently to an intensive study of AID gravity theories \cite{Biswas:2005qr,Biswas:2011ar} which were shown moreover to be easily made ghost free by adjusting the AID operators. This study was focused on a quadratic in $R$ Lagrangian. Note that absence of ghosts in such a setup can be achieved actually only by introducing an infinite number of derivatives, i.e. non-local operators. Further questions of renormalizability \cite{Modesto:2011kw}, presence of a ghost-free bounce \cite{Biswas:2012bp} and an amelioration of singularities \cite{Biswas:2011ar,Frolov:2015bta,Conroy:2016sac,Edholm:2016hbt,Koshelev:2017bxd} were addressed emphasizing in all instances a possibility to resolve successfully and consistently the problems in the framework of a quadratic in curvatures AID gravity modification (AID quadratic gravities/models/theories/etc. in short). On top of this ad-hoc AID scalar field models with a minimal coupling to GR were proven to be interesting in tackling the Dark Energy problem \cite{Aref'eva:2004vw,Arefeva:2006rnj,Arefeva:2008zru}.

There are several reasons to stick exactly with analytic differential operators. Primarily these are the presence of a well defined low-energy limit and their native appearance in a more fundamental approach which is SFT. However, as have been already mentioned the presence of exactly infinite number of derivatives (i.e. non-local operators) is a requirement to avoid ghosts. Interest in exactly quadratic in curvatures Lagrangians is stemmed from the fact that in many applications it is enough to study highly symmetric backgrounds, in particular Maximally Symmetric Space-times (MSS, which are in fact (Anti)-de Sitter ((A)dS) and Minkowski in the space-time dimension greater than 2) and linear perturbations around them. Linear perturbations in turn are described by the quadratic variation of the action.

It was proven by explicit construction in \cite{Biswas:2016etb}, that starting with a very generic action for the metric field such that the Lagrangian is analytic in curvatures and covariant derivatives, and focusing on the task of studying linear perturbations around MSS, one ends up with a quadratic in curvature action with analytic functions of the covariant d'Alembertian operator. In the most general case these analytic functions of derivatives become AID operators. No other combinations of derivatives apart from d'Alembertian and its AID functions appear. This is exactly the AID quadratic gravity and this is the most general and the only relevant Lagrangian we need to use in studying fluctuations around MSS.

The full gravity theory does not have to be just quadratic in curvatures. The point is that only the quadratic in curvatures part of some more general theory is responsible for the structure of propagators. This structure in turn in vast amount of situations determines whether the theory is unitary or not. However, another crucial property of a theory, its renormalizability, may require higher curvature terms present in the action \cite{Modesto:2015lna,Modesto:2014lga,Modesto:2017hzl}. Nevertheless the already observed properties of AID quadratic models make it inevitable to ask whether these theories are capable to eventually grow up to a full non-perturbative quantum theory of gravity. A significant step in this direction with a positive outcome was made recently in \cite{Koshelev:2017ebj} where super-renormalizable or finite quantum gravity candidates around MSS are constructed.

Given the success of AID quadratic gravity, it is natural to study whether it can admit inflationary solutions for some range of curvature. This is because cosmic inflation is not only a very successful theory of the early Universe \cite{Starobinsky:1980te,Guth:1980zm,Mukhanov:2005sc} but also at the same time for the moment is the best test-bed to challenge modified gravity theories. Viable models of inflation which can be parametrized by a number of free dimensionless parameters which values have to be fixed from observational data produce definite predictions about post-inflationary space-time metric perturbations given that an inflationary stage lasts long enough. The simplest models like the Starobinsky one have only one such parameter, so their predictive power is high. Further it explains the emergence of the Standard Model physics through the reheating mechanism \cite{Kofman:1994rk,Kofman:1997yn}.
However, in spite of a very large number of other inflationary models
already excluded by observations, still there remains a sufficient amount 
of them which remain viable, too, see e.g. \cite{Martin:2013nzq,Vennin:2015vfa}. In
many of these models having a larger number of free parameters, additional
scalar fields are introduced and gravity remains Einsteinian up to the
Planck curvature (unity in our notations). Even if we restrict ourselves,
as we do in this paper, to purely geometrical models of inflation and
modified gravity, still even in the $R^2$ model observations can probe a
large curvature regime only up to $R=4NM^2 \sim 240 M^2$ where $N$ is the
number of $e$-folds from the end of inflation back in time, that is much
less than unity.
In this regard AID gravity theories become natural candidates accounting the fact that they can be made ghost-free and tend to be renormalizable.

To have inflation is equivalent to guarantee a presence of solutions with a long enough nearly dS expansion and a subsequent graceful exit from this regime. To compute the key inflationary parameters one has to study linear perturbations around this nearly dS expansion phase. As we have explained above, AID quadratic gravity action is the maximal possible generalization one should ever consider for this purpose. This, as it was proven, covers considerations of inflation in arbitrary general original gravity theory as long as its action is analytic in curvatures and derivatives, and an appropriate inflationary solution exist. One equally can maintain a structural connection with other theories, like SFT, while this is not obviously necessary.

A first and successful try of embedding $R^2$ inflation into quadratic in $R$ AID gravity was performed in \cite{Craps:2014wga}. In a more recent paper \cite{Koshelev:2016xqb} it was argued that a particular quadratic in $R$ and in $W$ Lagrangian with AID operators is a renormalizable at least by power-counting and ghost-free gravity theory. The local $R^2$ inflation can be seamlessly embedded in this AID quadratic Lagrangian as well. Parameters of the new model allow to maintain a good agreement with the observational data at easy.

The main purpose of this paper is to deepen from the side of inflation the study of the AID quadratic gravity model undertaken in \cite{Koshelev:2016xqb}.
In what follows we will provide more support for the particular action used in that paper. The advance of the current analysis is the proof that the AID quadratic action which was derived in \cite{Biswas:2016etb} as the least non-redundant action for studying linear perturbations around MSS is in fact redundant thanks to Bianchi identities. Similar ideas were used in \cite{Biswas:2013kla} doing computations around the Minkowski background. Here we provide the full treatment around MSS and this is the purpose of Section~\ref{reduction}. Further we prove a very important statement that under certain assumptions the space of background space-homogeneous solutions in our AID model is identical to the space of backgrounds of a local $R^2$ gravity. This is a very important step since it allows to claim that the classical inflation remains an attractor behavior in the case of AID quadratic gravity. This is done in Section~\ref{classics0}. As the main accent of the present paper we systematically derive the inflationary parameters following from our model keeping the leading order in the slow-roll approximation throughout the whole computation. In particular we compute spectral tilts
 and tensor to scalar ratio. Note, that the previous studies assumed an exact dS background in the course of computation and applied the slow-roll approximation only starting from the action for canonical perturbation variable. The technique developed in this paper opens ways to restrict tighter the parameters of the new theory and to meet more and more toughly squeezed observational constraints. All the inflationary computations related to our model are accumulated in Section~\ref{perturbSec}. 
In Section~\ref{Concl} we discuss
the main results obtained in the paper and outline open questions. At last, extensive Appendices contain all the notations used in the paper as well as most technical pieces of the derivations.

\section{Most general AID quadratic gravity action around MSS}

\label{reduction}

One of the most crucial results of \cite{Biswas:2016etb} provides a most generic action for studying linear perturbations around MSS. Consider the following action 
(all notations without an explicit
explanation in the main text hereafter are accumulated in the Appendixes):
\begin{equation}
S=\intdg{4}\left[P_{0}+\sum_{i}\prod_{I}P_{i}O_{iI}Q_{iI}\right]\, ,\label{actionverygeneral}
\end{equation}
where $P,~Q$ depend only on the metric, Riemann tensor and curvatures
while $O$ depend only on covariant derivatives.
This action accommodates virtually all higher derivative gravity
theories with an analytic dependence on curvatures and covariant derivatives. Assuming an existence of at least one
(A)dS solution, the action relevant to study linear perturbations of EOM (coming from the quadratic variation of the action) around such a solution
boils down to 
\begin{eqnarray}
	\begin{split}
S&=\intdg{4}\left[\frac{M_{P}^{2}}{2}{R}\right.\\
&\left.+\frac{\lambda}{2}\left({R}\Fc_{1}(\Box){R}+R_{\mu\nu}\Fc_{2}(\Box)R^{\mu\nu}+R_{\mu\nu\rho\sigma}\Fc_{4}(\Box)R^{\mu\nu\rho\sigma}\right)-\Lambda\right]\,,\label{actionverygeneral2}
\end{split}
\end{eqnarray}
where $\lambda$ is a dimensionless constant which is convenient to
control the magnitude of the $R^{2}$ modification and $\Lambda$ is an
in principle possible cosmological constant term.
Briefly the reduction is
done by carefully accounting all possible terms which may contribute
non-trivially to the second variation around MSS (and dropping all
other terms). The fact which is heavily used on this way is that all
curvature tensors on MSS are annihilated by covariant derivatives.

An important assumption essential for the actual computations and which was discussed in the Introduction 
is that all functions $\Fc$ are analytic. To be precise we need at the moment to have these function analytic around zero. This is indeed required from the physical point of view. We want functions $\Fc(\Box)$ reduce to constants or vanish in a low-energy limit because we have to restore GR at very low energies. There is also other way to understand this. In writing $\Fc(\Box)$ we always assume that there is an energy scale of the gravity modification $\Mc$, which we name the scale of non-locality as in principle we may have infinite derivative operators ($\Mc$ should not be mixed with the much lower energy scale $M$ at which the $R^2/6M^2$  term in the local $R^2$ inflationary model becomes comparable to the GR term $R$). This scale enters as $\Fc(\Box/\Mc^2)$. Even though for most of our technical steps we can put $\Mc=1$ we still want to have a local or trivial limit once $\Mc\to\infty$ in order to be able to eventually restore GR. Hence, we come to the conclusion that functions $\Fc$ must be analytic at least in the origin. 

\textit{\textbf{Proposition:} Action (\ref{actionverygeneral2}) is redundant in describing linear fluctuations around MSS.}

This proposition can be proven to be true because the previous analysis did not make use of
Bianchi identities which is the cornerstone of the succeeding further
reduction. To start with, action (\ref{actionverygeneral2}) can be
rewritten as 
\begin{eqnarray}
	\begin{split}
	S&=\intdg{4}\left[\frac{M_{P}^{2}}{2}{R}\right.\\
&\left.+\frac{\lambda}{2}\left({R}\tilde{\Fc}_{R}(\Box){R}+L_{\mu\nu}\Fc_{L}(\Box)L^{\mu\nu}+W_{\mu\nu\rho\sigma}\tilde{\Fc}_{W}(\Box)W^{\mu\nu\rho\sigma}\right)-\Lambda\right]\,,\label{actionverygeneral2LC}
\end{split}
\end{eqnarray}
The purpose of using the Weyl tensor $W$ and $L$-tensor is
their simplicity. Both are identically zero on MSS. Moreover, $W$
is zero on any conformally flat background (including spatially flat FLRW).\footnote{Notice that in \cite{Biswas:2016etb} and \cite{Biswas:2016egy} we have used
$S_{\mu\nu}$ for what is now $L_{\mu\nu}$ and $C$ for Weyl tensor.
In the present paper we use $S$ for Schouten tensor and $C$ for
Cotton tensor for historical reasons.} The term to be attacked by Bianchi identities is the $L$-piece.
A good reason to tackle this term somehow is that being simple on
the background it produces tremendous complications while trying to
compute perturbations.

To make the long story short we put all the technical details of the reduction into Appendix~\ref{app:reduction}. Upon a lengthy but a straightforward procedure the full resulting action relevant for study of linear perturbations
around MSS vacua of (\ref{actionverygeneral}) becomes 
\begin{eqnarray}
S & = & \intdg{4}\left[\frac{M_{P}^{2}}{2}{R}+\frac{\lambda}{2}\left({R}\Fc_{R}(\Box){R}+W_{\mu\nu\rho\sigma}\Fc_{W}(\Box)W^{\mu\nu\rho\sigma}\right)-\Lambda\right]\,.\label{actionour4}
\end{eqnarray}
We consider this action as a significant simplification of (\ref{actionverygeneral2LC})
for several reasons:\footnote{
We note that our derivation is almost dimension independent. The only
local term which survives in higher dimensions is the local square of $L$-tensor
which we can drop in $D=4$ due to the presence of the Gauss-Bonnet
invariant. As such, the full action relevant for study of linear perturbations
around MSS vacua of (\ref{actionverygeneral}) formulated in $D>4$ can be written
as follows 
\begin{eqnarray}
S & = & \intdg{D}\left[\frac{M_{P}^{2}}{2}{R}+\frac{\lambda}{2}\left({R}\Fc_{R}\left(\Box\right){R}+L_{\mu\nu}^{2}+W_{\mu\nu\rho\sigma}\Fc_{W}(\Box)W^{\mu\nu\rho\sigma}\right)-\Lambda\right]\,.\label{actionourD}
\end{eqnarray}
We still consider using $L$-tensor is preferred as it is identically zero on
MSS.
}
\\
 (i) it contains only Ricci scalar and Weyl tensor and no Ricci tensor
or its linear combination with the metric. Weyl tensor enters only
quadratically and being identically zero on any conformally flat manifold
does not contribute to conformally flat background solutions. Importantly,
spatially flat FLRW metric is conformally flat.\\
 (ii) as such, any solution already found in the literature with only
$R\Fc_{R}(\Box)R$ piece in the action is a solution to equations
of motion which one can derive from our new action.\\
 (iii) linear perturbations of Weyl tensor are very simple using $(1+3)$
decomposition of the ADM formalism. These were computed in \cite{Koshelev:2016xqb}
and one can track computations relevant to our AID models in
application to inflation to the end. Actually, perturbations of a possible term with any of the second
rank tensors (Ricci, Schouten, Einstein or $L$-tensor) turn out to
be very much complicated and seem to be intractable.

It  is worth stressing that actions (\ref{actionverygeneral2}) and (\ref{actionverygeneral2LC}) are not fully equivalent. They are equivalent as long as at most linear perturbations around MSS are considered. As a consequence, non-MSS may be solutions to EOM derived from one action and not from another. For example, local $R^2$ inflationary background is a solution to EOM derived from action (\ref{actionverygeneral2LC}) and is not a solution as long as the quadratic term with a second-rank tensor is restored in the action. Furthermore, higher, i.e. 3-, 4-, \dots-point vertices and correlation functions are clearly different in these actions. It is however of course possible that the difference may not be important for particular models and under certain assumptions.

\section{Classical dynamics of AID quadratic gravity without $\Lambda$}
\label{classics0}
\subsection{EOM and solution construction}

The major focus of the present paper is on the consideration of inflation in AID quadratic gravity and for that purpose the cosmological term is not needed. It is shown already in \cite{Starobinsky:1980te} that in fact a cosmological term spoils the inflation. To proceed with actual computation we recite action (\ref{actionour4}) dropping the cosmological term $\Lambda$:
\begin{eqnarray}
S & = & \intdg{4}\left[\frac{M_{P}^{2}}{2}{R}+\frac{\lambda}{2}\left({R}\Fc_{R}(\Box){R}+W_{\mu\nu\rho\sigma}\Fc_{W}(\Box)W^{\mu\nu\rho\sigma}\right)\right]\,.\label{actionour4inf}
\end{eqnarray}
This action was studied and many technical details were elaborated in \cite{Biswas:2012bp,Craps:2014wga,Koshelev:2016xqb}. We are going to use them without extensive further referencing.

EOM which one can derive from action (\ref{actionour4inf}) (see \cite{Koshelev:2013lfm}) read:
\begin{eqnarray}
	E^\mu_\nu &\equiv& -(M_P^2+2\lambda\Fc(\Box)R)G^\mu_\nu-\nonumber
	\frac12\lambda
	 {\delta}^\mu_\nu R\Fc(\Box)
	 R+2\lambda(\nabla^\mu\partial_\nu-\delta^\mu_{\nu}
	 \Box) \Fc(\Box) R
	 \nonumber\\
	 &+&\lambda{\Lc}^\mu_\nu
	 -\frac{\lambda}{2}\delta^\mu_{\nu}\left({\Lc}^\sigma_\sigma+\bar\Lc
	 \right)+
	 2\lambda
	 \left(R_{\alpha\beta}
	 +2\nabla_\alpha\nabla_\beta\right)
	 \Fc_{W}(\Box)W_{\nu}^{\phantom{\nu}\alpha\beta\mu}
	 +\mathcal{O}({W}^2)=0 \, . 
	 \label{tEOM}\\
	 {\Lc}^\mu_\nu&=& \sum_{n=1}^\infty
	 {{ f}}{}_n\sum_{l=0}^{n-1}\partial^\mu  R^{(l)}  \partial_\nu  R^{(n-l-1)} \, , \quad 
	 \bar\Lc=\sum_{n=1}
 ^\infty
 {f}{}_n\sum_{l=0}^{n-1} R^{(l)}    R^{(n-l)}\, ,\quad R^{(l)} \equiv \Box^l R \, ,\nonumber
 \end{eqnarray}
The trace equation reads
\begin{equation}
	E = (M_P^2-6\lambda\Box\Fc_R(\Box))R-\lambda(\Lc^\mu_\mu+2\bar\Lc)+\mathcal{O}(W^2)=0
	\label{tEOMtrace}
\end{equation}
Terms linear in Weyl tensor are not present in the trace equation because their trace vanishes by construction on any space-time. Terms $\mathcal{O}(W^2)$ can be found in \cite{Biswas:2013cha}.

We are interested in cosmological solutions of the spatially flat FLRW type. First this implies that the Weyl tensor vanishes and as such it does not manifest itself in the trace equation neither in the background nor in linear perturbations. Second, such solutions for the metric are space-homogeneous and isotropic.
This means that system of equation (\ref{tEOM}) has essentially two distinct equations. The standard choice is the trace equation and the $({}^0_0)$-equation. However, presence of Bianchi identities guarantees that given we have a solution to the trace equation with zero RHS then it will be a solution to the whole system of equations modulo a possible radiation source (which is conserved and is traceless). We are thus focused on solving the trace equation (\ref{tEOMtrace}) which is a non-linear differential (non-local) equation on $R$ and all the differential operators are of the form of d'Alembertian.

We start solving the trace equation by reminding that originally it was proposed in \cite{Biswas:2005qr} to use an ansatz
\begin{equation}
	\Box R=r_1 R
	\label{wasans}
\end{equation}
to construct solutions. First we note that the original ansatz also had a free constant term $r_2$ in the right hand side but it is not compatible with the absence of the cosmological term. Also we note that this ansatz was indeed helpful to construct several exact solutions to equation of motion.

It is instructive to show sketchy how the technique works. Substituting (\ref{wasans}) into (\ref{tEOMtrace}) we restore the result obtained in \cite{Biswas:2005qr}
\begin{equation}
	(M_P^2-6\lambda r_1\Fc_R(r_1))R-\lambda\Fc_R^{(1)}(r_1)(\pd^\mu R\pd_\mu R+2r_1R^2)=0
	\label{tEOMtrace1}
\end{equation}
The way to solve the latter equation is to assume the algebraic conditions
\begin{equation}
	\Fc^{(1)}_R(r_1)=0,~\frac{M_{P}^{2}}{2\lambda}=3r_{1}\mathcal{F}_{1},~\text{ where }\Fc_1\equiv\Fc_R(r_1)
	\label{conditions}
\end{equation}
Since we constrain here only parameters we get the most what we can using (\ref{wasans}). If we do not impose the above conditions then we must satisfy additional conditions on $R$ which can be shown to trivialize possible solutions to just one $R=0$. We accumulate the details supporting this claim in Appendix~\ref{wasansdiff:app}. This will become useful in the coming Subsection.

\subsection{Proof that (\ref{wasans}) is general solution to (\ref{tEOMtrace})}\label{sec:proposition}

Now we formulate the main claim of this Section and in fact a very important statement for the development of AID quadratic gravity theories in general.

\textit{\textbf{Proposition:} equation (\ref{wasans}) in combination with conditions (\ref{conditions}) provide the most general solution to the trace equation (\ref{tEOMtrace}) if:\\
(i)~the metric is of a spatially flat FLRW type and\\
(ii)~the Fourier harmonics form a basis on the domain of functions of interest on the space-time manifold.}

Let us start with noting that having a physical attitude to the problem we formulate here sufficient and not obligatory necessary conditions.

The first condition \textit{(i)} just serves for the setting of the present paper to discuss a space-homogeneous inflationary space-time and is simple to account. Technically during the proof the only property to be exploited will be the space-homogeneity of the metric in the synchronous frame. As such the proof itself can be applied to more general space-homogeneous metrics. For example to anisotropic backgrounds like Bianchi~I or other. However, we need the metric to be conformally flat to eliminate the Weyl tensor squared terms in the trace equation (\ref{tEOMtrace}). This would allow us to claim that the restrictions imposed by this proposition on the space of solutions apply to the full trace equation. Only for this purpose we stick to spatially flat FLRW metrics only. This implies that
given the Weyl tensor dependent term is not included in the action (\ref{actionour4inf}) one can relax condition \textit{(i)} to:
\textit{(i)~the metric is space-homogeneous in the synchronous frame.}

The second condition \textit{(ii)} needs more explanations though. We name Fourier harmonics the eigenfunctions of the d'Alembertian operator such that
\begin{equation}
	\Box \varphi_i=w_i\varphi_i
	\label{efbox}
\end{equation}
where $w_i$ are constants. Generically we expect the spectrum of the d'Alembertian is continuous even though this is not crucial. We name $\varphi_i$ the Fourier harmonics in analogy with the flat space-time where they reduce to the plane waves which in turn are used do define the Fourier transform. A crucial property of the Fourier transform in the flat space-time is that the corresponding harmonics form a basis in the domain of square integrable functions $L_2$. Or in other words any square integrable function can be presented as a linear superposition of plane waves. In our model the situation seems to be more involved as a priori these nice properties of the Fourier transform in the flat space-time cannot be elevated to a curved background.

It is known from the spectral analysis of the Beltrami-Laplace (BI) operator on Riemannian manifolds that indeed the eigenmodes of the BI operator form a basis in $L_2$ as long as the manifold is compact or has a boundary \cite{helgason2001differential}. In most cosmological applications the space-time manifold is however pseudo-Riemannian (i.e. the metric is not positively defined and d'Alembertian operator replaces the BI operator), non-compact and without a boundary. In this situation general theorems do not help and presently one has to consider systems case-by-case. Paper \cite{Balasubramanian:2002zh} provides an explicit proof that in two notable cases of dS and (A)dS space-times indeed the eigenmodes of the d'Alembertian operator form a basis for square integrable functions.
This remains valid in a special situation when there is no spatial dependence is present. It is an important situation though since in the vast majority cosmologically viable backgrounds are space-homogeneous. Naturally, it is the case for the present paper as well. Technically, this implies that the d'Alembertian operator lacks of spatial derivatives and eigenmodes $\varphi_i$ depend on time only.

Coming to physical grounds we stress that the regime of the space-time evolution of interest in the present paper is the nearly dS expansion. This in combination with the results in \cite{Balasubramanian:2002zh} provides some hint that our condition \textit{(ii)} in the proposition above is sensible. However, there is one more physically important argument why a physically viable space-time must have such a structure that the d'Alembertian operator eigenmodes form a basis. Namely, we expect that our model can be quantized. To have this happen we have to have a vacuum \textit{and} creation and annihilation operators which in the canonical quantization scheme appear as operator coefficient in front of Fourier modes in which a given classical solution is decomposed. Given a situation that Fourier modes do not form a basis (i.e. the set of modes is not enough to represent any function) we will hit a problem that certain classical configurations cannot be quantized in a canonical way. This simple consideration shows that the fact that eigenmodes of the d'Alembertian operator form a basis is necessary to implement the canonical quantization scheme. This gives us even a stronger hint that we indeed want the condition \textit{(ii)} in the proposition to be satisfied.

Finally, we do not specify explicitly the domain of functions on which the completeness of the Fourier decomposition is true. We presume that in most cases we need to have it either for functions from $L_2$ or functions with a compact support which is a more plausible case as long as time evolution of a classical system is considered. This will be noted just below as well.

Therefore, in proving the proposition we assume that the scalar curvature $R$ as any function can be represented as
\begin{equation}
	R=\sum_i \varphi_i\, ,\qquad\Box \varphi_i=w_i\varphi_i
	\label{ourans}
\end{equation}
and $w_i$ are constants. Possible constants in front of $\varphi_i$ in the decomposition of $R$ are absorbed inside of $\varphi_i$ for simplicity. Few comments are in order here. First, one should not be confused with the fact that $R$ itself depends on the metric as for the time being it is just some function of time. Second, one should not worry about possible non-trivial asymptotics of $R$ in past or future infinities (which may render it non-square integrable) and consider only a given time interval during which our model describes the evolution of the Universe. This will waive doubts of the square integrability since the function gains the compact support by construction. In other words it is equivalent to saying that we work in a given coordinate patch. Also, here we explicitly come to the special situation mentioned above that all functions depend on time only since a space-homogeneous background is considered. The corresponding simplification will become crucial to fulfill the proof.

Using (\ref{ourans}) one readily computes
\begin{equation}
	\Box^l R=\sum_i w_i^l\varphi_i,\quad \Fc_R(\Box)R=\sum_i\Fc_R(w_i)\varphi_i
	\label{ouransrecursion}
\end{equation}
and further
\begin{equation}
	\Lc^\mu_\mu=\sum_{i,j}\omega_{ij}\pd^\mu \varphi_i\pd_\mu \varphi_j,\quad\bar\Lc=\sum_{i,j}\omega_{ij}w_j\varphi_i\varphi_j,\quad\omega_{ij}=\frac{\Fc_R(w_i)-\Fc_R(w_j)}{w_i-w_j}
	\label{Lrecursion}
\end{equation}
Notice that for $i=j$ we have using the Taylor series expansion $\omega_{ii}=\Fc_R^{(1)}(w_i)$ where the superscript ${}^{(1)}$ denotes the derivative with respect to an argument.
Substituting all of that into (\ref{tEOMtrace}) and accounting that the Weyl tensor vanishes one yields
\begin{equation}
	M_P^2\sum_k\varphi_k-6\lambda \sum_kw_k\Fc_R(w_k)\varphi_k-\lambda\sum_{i,j}\omega_{ij}(\pd^\mu \varphi_i\pd_\mu \varphi_j+(w_i+w_j)\varphi_i\varphi_j)=0
	\label{tEOMtracerecursion}
\end{equation}
To prove the proposition we have to show that no (non-trivial) solutions to (\ref{tEOMtracerecursion}) exist as long as $R$ is a superposition of more than a single Fourier eigenmode.

First we note that the technique of equating coefficient to zero does not work in this general case. Indeed, the quadratic in $\varphi_i$ term in (\ref{tEOMtracerecursion}) can be eliminated by requiring $\Fc_R(w_i)=\Fc_R(w_j)$ and $\Fc_R^{(1)}(w_i)=0$ for any $i,j$. This being substituted into the terms linear in $\varphi_i$ yields
\begin{equation*}
	M_P^2\sum_k\varphi_k-6\lambda \Fc_R(w_1)\sum_kw_k\varphi_k=0
\end{equation*}
Since however different $\varphi_k$ are eigenfunctions of d'Alembertian with different eigenvalues they are linearly independent. This means that in order to satisfy the latter equation we must require $M_P^2-6\lambda \Fc_R(w_1)w_k=0$ for each $k$ and as such all $w_k$ are equal. We thus effectively come back to the situation $R=\varphi_1$ like it is served by (\ref{wasans}).

Thus we must keep the quadratic terms in (\ref{tEOMtracerecursion}) and solve it as a differential equation on $\varphi_i$. Satisfying (\ref{tEOMtracerecursion}) will necessarily produce stringent constraints since the resulting solution for $R$ must be identical to the Ricci scalar constructed from the metric. Note, that in the beginning of the proof we have mentioned that $R$ is just some function of time. Here we explicitly make reference to its relation to the metric. This, however, in no way complicates the use of desired spectral properties of the d'Alembertian.

Going further one can pass to modified quantities $\tilde \varphi_i=\varphi_i+c_i$ where we have done shifts by constants defined as
\begin{equation}
	2\lambda\sum_j \omega_{kj}(w_k+w_j)c_j+M_P^2-6\lambda w_k\Fc_R(w_k)=0\text{ for each }k
\end{equation}
Hence we rewrite (\ref{tEOMtracerecursion}) as
\begin{equation}
	\sum_{i,j}\omega_{ij}\dot{\tilde \varphi}_i\dot{\tilde \varphi}_j-\sum_{i,j}\omega_{ij}(w_i+w_j)\tilde \varphi_i\tilde \varphi_j=c=-\sum_{i,j}\omega_{ij}(w_i+w_j)c_ic_j
	\label{tracedyn}
\end{equation}
where we have used the fact that all $\varphi_i$ are space-homogeneous and depend only on time. The sign change in front of $\sim \dot{\tilde \varphi}_i^2$ is due to the signature of the metric and also we assume that $g_{00}=-1$. Also a common factor $\lambda$ has been cancelled.
Interestingly, we recognize in the latter formula the conserved integral of energy originating from a sigma-model-type dynamical system.

To make the succeeding analysis more transparent we rewrite the last formula using matrix notations as follows
\begin{equation}
	\dot{\widetilde {\boldsymbol{R}}}^{\mathrm{T}}{\boldsymbol{\omega}}\dot{\widetilde{\boldsymbol{R}}}-\widetilde{\boldsymbol{R}}^{\mathrm{T}}(\boldsymbol{w\omega}+\boldsymbol{\omega w})\widetilde{\boldsymbol R}=c
	\label{tracematrix}
\end{equation}
where $\widetilde{\boldsymbol{R}}$ is a vector made of $\tilde\varphi_i$ and ${\boldsymbol{w}}=\mathrm{diag}({w_1,w_2,\dots})$ and $\boldsymbol{\omega}$ is a matrix formed by $\omega_{ij}$. We use a simple transposition as all quantities are real and matrices are symmetric from the physical origin of the problem. We diagonalize matrix $\boldsymbol{\omega}$ by choosing an appropriate matrix $\boldsymbol{D}$. We can always do this because if $\boldsymbol{\omega}$ cannot be diagonalized then some values $w_i$ are identical and we must just drop equivalent terms from decomposition (\ref{ourans}). Denoting ${\boldsymbol{d}}^2={\boldsymbol{D}}^{\mathrm{T}}\boldsymbol{\omega D}$ and using further redefined functions $\boldsymbol{Q}={\boldsymbol{d D}}^{\mathrm{T}}\widetilde{\boldsymbol{R}}$ we get a canonically normalized diagonal term with derivatives. The whole expression transforms as
\begin{equation}
	\dot{{\boldsymbol{Q}}}^{\mathrm{T}}\dot{{\boldsymbol{Q}}}-{\boldsymbol{Q}}^{\mathrm{T}}{\boldsymbol{\nu}}{\boldsymbol Q}=c
	\label{tracemq}
\end{equation}
where $\boldsymbol{\nu}={\boldsymbol{d}}^{-1}{\boldsymbol{D}}^{\mathrm{T}}\boldsymbol{w D d}+\boldsymbol{d}{\boldsymbol{D}}^{\mathrm{T}}\boldsymbol{w D}{\boldsymbol{d}}^{-1}$. We can simplify the things even more by diagonalizing the matrix $\boldsymbol{\nu}$ choosing an appropriate matrix $\boldsymbol{M}$. Denoting $\boldsymbol{m}^2=\boldsymbol{M}^{\mathrm{T}}\boldsymbol{\nu M}$ and redefining $\boldsymbol{P}=\boldsymbol{M}^{\mathrm{T}}\boldsymbol{Q}$ we get
\begin{equation}
	\dot{{\boldsymbol{P}}}^{\mathrm{T}}\dot{{\boldsymbol{P}}}-{\boldsymbol{P}}^{\mathrm{T}}{\boldsymbol{m}^2}{\boldsymbol P}=c
	\label{tracemp}
\end{equation}
Here the most crucial achievement that matrix $\boldsymbol{m}$ is diagonal.

Differentiating the latter equation with respect to the time $t$ one gets
\begin{equation}
	\dot{{\boldsymbol{P}}}^{\mathrm{T}}(\ddot{{\boldsymbol{P}}}-{\boldsymbol{m}^2}{\boldsymbol P})=0
	\label{eqp}
\end{equation}
As noticed above, all $\varphi_i$ are linearly independent and the same are $P_i$. Indeed, since the matrices which define the quadratic form are non-degenerate this guarantees that $P_i$ are linearly independent. We thus can consider only the second order linear equations in the latter expression as all of them must be satisfied independently. Since moreover matrix $\boldsymbol{m}$ is diagonal we readily find each $P_i$ as
\begin{equation}
	P_i=P_{i+}e^{m_it}+P_{i-}e^{-m_it}\, ,
	\label{eqpsol}
\end{equation}
where we have assumed $\boldsymbol{m}=\mathrm{diag}(m_1,m_2,\dots)$.

Returning to (\ref{ourans}) we can rewrite the corresponding expression in matrix notations as well.
That is
\begin{equation}
	\ddot{\boldsymbol{R}}+3H\dot{\boldsymbol{R}}+\boldsymbol{wR}=0\, .
	\label{ouransmx}
\end{equation}
Note that the latter equation is valid for any space-homogeneous metric as long as $g_{00}=-1$ and as such $H$ is the Hubble function only in the case of a spatially flat FLRW metric.
Passing to variables $P_i$ we get
\begin{equation}
	\ddot{\boldsymbol{P}}+3H\dot{\boldsymbol{P}}+\boldsymbol{\mu P}=\boldsymbol{\chi c}
	\label{ouransmxP}
\end{equation}
where $\boldsymbol{\mu}=\boldsymbol{M}^{\mathrm{T}}\boldsymbol{d}{\boldsymbol{D}}^{\mathrm{T}}\boldsymbol{w D}{\boldsymbol{d}}^{-1}\boldsymbol{M}$, $\boldsymbol{\chi}=\boldsymbol{M}^{\mathrm{T}}\boldsymbol{d}{\boldsymbol{D}}^{\mathrm{T}}\boldsymbol{w}$ and $\boldsymbol{c}=\mathrm{diag}(c_1,c_2,\dots)$.

To prove the proposition we must show that solutions (\ref{eqpsol}) are incompatible with (\ref{ouransmxP}) for more than one component vector $\boldsymbol{P}$. Being lucky that we could construct solutions for $P_i$ explicitly we just substitute them into (\ref{ouransmxP}). The resulting expression is
\begin{equation}
	\boldsymbol{m}^2{\boldsymbol{P}}+3H\boldsymbol{m}(\boldsymbol{P}_+-\boldsymbol{P}_-)+\boldsymbol{\mu P}=\boldsymbol{\chi c}
	\label{ouransmxPsol}
\end{equation}
where $\boldsymbol{P}_\pm=\mathrm{diag}(P_{i\pm}e^{\pm m_it})$. Each component $P_i$ is a different exponent and in order to satisfy the latter equation we must put to zero coefficients in front of each of them. Moreover, we must have the constant term on the right hand side vanish.
If $H\neq0$ we essentially must require the matrix $\boldsymbol{m}$ to be zero and this is equivalent to
having all $w_i=0$ and as such we come back to the situation $\Box R=0$ which is just a sub-case of (\ref{wasans}) and in no way requires any more general form of $R$ then a single Fourier mode.

This completes the proof of the proposition during which we actually have never used that the metric must be exactly of a spatially flat FLRW type.

In a slightly exotic situation such that there is a space-homogeneous metric which generates vanishing factor $H$ in (\ref{ouransmx}) in a combination with a non-constant $R$ one need to have a separate consideration regarding the space of solutions to EOM.

\subsection{Discussion on classical dynamics}

We just have proven a very important fact related to theories of type (\ref{actionour4inf}): all space-homogeneous conformally flat background solutions are subject to equation (\ref{wasans}) in combination with conditions (\ref{conditions}).

To understand what happens we must examine conditions (\ref{conditions}) which tell that a non-trivial solution (i.e. a solution more involved than a constant $R$) exists only if there is a point $r_1$ such that function $\Fc_R(r_1)$ being considered as a function of $r_1$ as its parameter obeys two independent algebraic conditions. Moreover, a would be solution must obey an equation which can be derived from a local $R^2$ gravity. It was elaborated in \cite{Koshelev:2016xqb} what a Lagrangian for a local model must be written such that its equation of motion yields $\Box R=r_1 R$. So essentially we should worry whether function $\Fc_R(r_1)$ provides a chance to have at least some solution.

The other point of view is to consider presence of a solution as a criterion for function $\Fc_R(r_1)$ such that it provides a choice of points $r_1$ at which conditions (\ref{conditions}) are true. Since functions $\Fc_X(\Box)$ are not constrained so far apart from being analytic at the origin one may wonder about the space of solutions. Indeed, it is possible that a generic function $\Fc_R(r_1)$ has many or may be even infinitely many points in which $r_1\Fc_1$ is the same value and plus to this $\Fc_R^{(1)}(r_1)=0$ at these points. This is some sense would mean that our model include multiple copies of local $R^2$ gravity.

Even though mathematically possible we are going to remind that the operator functions $\Fc_X(\Box)$ get severely constrained by demand that no new excitations must appear in the spectrum.
Indeed, as it was derived in all the details in \cite{Biswas:2016egy} the quadratic Lagrangian for the spin-0 mode of the metric around the Minkowski space-time (where we must fix the operator functions) is
\begin{equation}
	S_0=-\frac12\int d^4x\sqrt{-\bar g}\phi\bar\Box\left\{1-2\frac{\lambda}{M_P^2}3\bar\Box\Fc(\bar\Box)\right\}\phi
	\label{S20}
\end{equation}
In order to contain the spectrum of excitations and to have inflation we must require that the expression in curly brackets has exactly one zero which would correspond to the scalaron. This can be achieved by demanding
\begin{equation}
	1-2\frac{\lambda}{M_P^2}3\bar\Box\Fc(\bar\Box)=\sigma_0(\bar \Box-M^2)e^{2\sigma(\bar\Box)}
	\label{S20scalaron}
\end{equation}
Here $\sigma(\Box)$ must be an entire function. For the definiteness we assume that $\sigma(0)=0$ and in order not to lose generality we introduce $\sigma_0$. Recall that $\Fc(0)=f_0$ and evaluating left and right sides of (\ref{S20scalaron}) at $\bar\Box=0$ we get
$$1=-\sigma_0 M^2$$
yielding $\sigma_0=-1/M^2$. Next, evaluating (\ref{S20scalaron}) at $\bar\Box=r_1$ and accounting (\ref{conditions}) we get
$$0=-\frac1{M^2}(r_1-M^2)e^{2\sigma(r_1)}$$
This implies that $r_1=M^2$.
Differentiating (\ref{S20scalaron}) with respect to the d'Alembertian, evaluating the result at $\bar\Box=r_1$ and accounting (\ref{conditions}) one gets
$$-2\frac{\lambda}{M_P^2}3\Fc_1=-\frac1{M^2}e^{2\sigma(r_1)}$$
This implies $\sigma(r_1)=0$.

The above results confirm the derivation done in \cite{Koshelev:2016xqb}. However the more important observation is that the condition $r_1=M^2$ together with the demand that only one excitation can exist guarantees that from the point of view of the physics of our model only a single unique point $r_1$ is allowed.

This is a very powerful statement because it implies that as long as space-homogeneous and conformally flat metrics are considered our quadratic AID gravity has exactly the same space of solutions as a local $R^2$ gravity. In particular this means that the inflationary background will remain an attractor behaviour as it was found originally for the local $R^2$ model in \cite{Starobinsky:1980te}.

To conclude this Section we notice that there are already mentioned limitations of our analysis: we are talking about space-homogeneous and conformally flat solutions (while allowing anisotropic metrics like of the Bianchi I types in the absence of the Weyl tensor term from the very beginning), we do not have matter sources apart from perhaps radiation.
Also, we do not have the cosmological term in the action. It was mentioned above that the presence of a cosmological term requires (\ref{wasans}) to be modified as follows
\begin{equation}
	\Box R=r_1R+r_2
	\label{wasanscc}
\end{equation}
In this case, for instance, the whole proof of the proposition from the previous Subsection must be reconsidered if at all possible.
We keep the generalization of our analysis to the models with the cosmological term as well as other interesting directions of further developments for future projects.

\section{Universal inflationary predictions from general gravity}

\label{perturbSec}

\subsection{General considerations}

As it was understood in Section~\ref{reduction} any arbitrary analytic in curvatures and derivatives gravity action (\ref{actionverygeneral}) is governed around MSS in $D=4$ by action (\ref{actionour4}) or (\ref{actionour4inf}) if no cosmological term is present. We note that the cosmological term is not needed at all for the inflation to happen. Inflation is the nearly dS expansion phase and thus action (\ref{actionour4inf}) is the correct starting point to compute inflationary parameters. Therefore having computed and confronted with measurements parameters of inflation from action (\ref{actionour4inf}) one can further try to approach the full theory of gravity.

As it was further crucially proven in Sections~\ref{classics0}, given one seeks for space-homogeneous and conformally flat metrics, our theory features only solutions which can be found in a local $R^2$ gravity. Technically this means that any solution must obey equation (\ref{wasans}) and due to the ghost-free conditions only one non-zero $r_1$ parameter is allowed.
Hence, the local inflationary background is seamlessly embedded in our most general consideration. We stress that this particular background is the minimal choice for the physically viable cosmological inflation which must be a long enough nearly dS phase of expansion with a graceful exit. Thanks to the proof of absence of other solutions in our model we guarantee that the local $R^2$ inflation remains an attractor solution which is an important property from the physical point of view.

Surely, perturbations are expected to be different from a local gravity. Since the main results in consideration of any inflationary model are related to perturbations we focus on analyzing them as detailed as possible and this is the main focus of this Section.
 Note that the best done so far computations of inflationary parameters \cite{Koshelev:2016xqb,Edholm:2016seu} treated the background as exact dS until the action for canonical perturbation variables.
 In the present analysis we are going to maintain consistently the leading slow-roll approximation throughout all the computations.
To control the slow-roll approximation we use the pretty much standard slow-roll parameter $\epsilon$ defined as
\begin{equation}
\epsilon=-\frac{\dot{H}}{H^{2}}=1-\frac{\mathcal{H}^{\prime}}{\mathcal{H}^{2}}\,.
\label{slowroll}
\end{equation}
Its use is elaborated in Appendix~\ref{slowroll:app}.

Thus, technically we should work with the solution to equation (\ref{wasans}) which was solved already in \cite{Starobinsky:1980te}.
The inflationary phase of this solution is known approximately \cite{Starobinsky:1980te,Koshelev:2016xqb}
\begin{eqnarray}
 a(t) &\approx& a_0 ({t_s - t})^{-1/6} \, e^{-r_1 (t_s -t)^2/12} \, ,
	\label{starinffactor}
\label{starR}
\end{eqnarray}
and one can check that the slow-roll approximation is valid for it for appropriate choice of parameters. Here $t_s$ is the time of the end of inflation.
Also, in principle, the solution to the whole system of Einstein equations may require radiation $\rho_r$. Its amount is given by (see \cite{Koshelev:2016xqb}):
\begin{equation}
	\frac{M_P^2}{3M^2}\left[\frac32 R\dot H-3H\dot R-9M^2H^2\right]+\rho_r=0 \, ,
	\label{00radnonlocal}
\end{equation}
and in the leading slow-roll approximation the radiation source vanishes (see Appendix~\ref{slowroll:app}).
The latter consideration of energy density of radiation emphasizes that given we do not consider its perturbations we are limited by the linear order in the slow-roll parameter in our analysis as any further expansion would require to include perturbations of the radiation fluid into consideration.

\subsection{Action for perturbations}

In order to analyze perturbations and their properties one can either analyze the linear variation of EOM or an action for perturbations which is the second order variation of the background action. While the results must be the same irrespectively of the approach some steps may be more simple in either of them. Linearization of EOM worked well in previous papers but a construction of the second order variation of the action was possible only around MSS.

Below we present for the first time the second order variation of (\ref{actionour4inf}) around any solution satisfying (\ref{wasans}) and conditions (\ref{conditions}). To do so we introduce an auxiliary local action
\begin{equation}
	S_{local}=\intdg{4}\left(\frac {M_P^2}2R+\frac\lambda 2\Fc_1R^2\right)\, .
	\label{actionour4inflocal}
\end{equation}
The answer for the second variation of (\ref{actionour4inf}) turns out to be unexpectedly simple and it reads after some laborious steps outlined in Appendix~\ref{s2:app}
\begin{equation}
	\delta^2S=\intdg{4}\left[\delta_{local}+\frac\lambda2\zeta \Zc_2(\bar\Box)\zeta+\frac\lambda2(\delta W)\Fc_W(\bar\Box)\delta W\right]\, ,
	\label{delta2Sinf}
\end{equation}
where
\begin{equation*}
	\delta_{local}=\frac{\lambda\Fc_1}2\left[2(\bar R+3r_1)\delta_0-\bar R^2\delta_g+(\delta R)^2\right]
\end{equation*}
is exactly the second order variation coming from a local action (\ref{actionour4inflocal}). Quantities $\delta_0=(\delta^{2}\sqrt{-g}R)/\sqrt{-g}$ and $\delta_g=(\delta^{2}\sqrt{-g})/\sqrt{-g}$ are computed explicitly in Appendix~\ref{s2:app}. Further,
\begin{equation}
	\zeta=(\bar\Box-r_1)\delta R+(\delta\Box) \bar R\, .
	\label{zetamain}
\end{equation}
This is essentially a variation of (\ref{wasans}) and it would be identically zero in a local case but may not be assumed as such as (\ref{wasans}) is not an EOM in our AID action. $\zeta$ actually becomes an essential quantity in our model. Finally,
\begin{equation}
	\Zc_2(\Box)=\frac{\Fc_R(\Box)-\Fc_1}{(\Box-r_1)^2}\label{z2main}
\end{equation}
is an operator analytic in $\bar\Box$ as can be shown using the Taylor series expansion in combination with (\ref{conditions}). As usual quantities with bars over them are the background ones.

First we note that the latter action for perturbations is derived without any assumptions on the background apart from the fact that it satisfies (\ref{wasans}) and conditions (\ref{conditions}) are satisfied. It is a general action valid in all regimes and not only around the actual dS expansion of the Universe. This is a complete expression for all modes: scalars, vectors and tensors. The action differs from a known answer for a local $R^2$ inflationary model by just one term square in $\zeta$. This term is actually non-local due to the presence of the operator $\Zc_2(\Box)$. The last term containing the Weyl tensor variations is that simple because the Weyl tensor is zero on the backgrounds of interest and as such nothing else can survive upon the second variation.

We proceed by considering scalar and tensor perturbations meaning that the classification is with respect to the 3-dimensional symmetry group.

\subsection{Scalar perturbations}
The perturbed line element for scalar perturbations in terms of Bardeen
potentials $\left(\Phi,\,\Psi\right)$ reads
\begin{equation}
ds^{2}=a^{2}\left(\tau\right)\left[-\left(1+2\Phi\right)d\eta^{2}+\left(\left(1-2\Psi\right)\delta_{ij}+2h_{ij}\right)dx^{i}dx^{j}\right]\,.\label{line-element}
\end{equation}
The gauge invariant perturbation of the scalar curvature is given by 
\begin{equation}
\delta R_{GI}=2\left(\bar{R}+3\bar{\Box}_{k}\right)\Psi-2\bar{R}\left(\Phi+\Psi\right)-6\frac{\mathcal{H}}{a^{2}}\left(\Phi^{\prime}+\Psi^{\prime}\right)+2\frac{k^{2}}{a^{2}}\left(\Phi+\Psi\right)\,.\label{deltaRGI}
\end{equation}
This is used instead of $\delta R$ in the definition of $\zeta$ in (\ref{zetamain}) as long as we pass to gauge invariant variables. As it is shown above $\zeta$ essentially measures the difference of our model from the local $R^2$ gravity (see action (\ref{delta2Sinf})). Also we recall that being a variation of (\ref{wasans}) it is zero in a local $R^2$ gravity but is not obligatory trivial in our case.

However, $\zeta$ is governed by a linear and homogeneous though non-local equation (\ref{dE}) which we recite here for the completeness
\begin{equation}
\left\{\left[\partial^{\mu}\bar{R}\partial_{\mu}+2r_{1}\bar{R}\right]\mathcal{Z}_{2}\left(\bar{\Box}_{k}\right)+
3\mathcal{\mathcal{F}}_{R}\left(\bar\Box_k\right)+\left(\bar{R}+3r_{1}\right)\mathcal{Z}_{1}\left(\bar\Box_k\right)\right\}\zeta=0\, ,
\label{dEtext}
\end{equation}
where
\begin{equation}
	\begin{aligned}
\mathcal{Z}_{1}\left(\Box\right)=  \frac{\mathcal{F}_{R}\left(\Box\right)-\mathcal{F}_{1}}{\left(\Box-r_{1}\right)}\,,\quad
\mathcal{Z}_{2}\left(\Box\right)=  \frac{\mathcal{F}_{R}\left(\Box\right)-\mathcal{F}_{1}}{\left(\Box-r_{1}\right)^{2}}\,.
\end{aligned}
\label{z1z2}
\end{equation}
We are going to explore the solutions to the above equation in the leading slow-roll approximation. This generalizes our consideration in \cite{Koshelev:2016xqb} because there we have taken the pure dS background for inflation which is the zero level approximation. Strictly speaking however, having the background (\ref{starinffactor}) up to the leading slow-roll correction we have to follow the same approximation in computing perturbations.

Using the details accumulated in Appendix~\ref{app:linear} we state the essence of this consideration that in the leading slow-roll approximation the only possibility for $\zeta$ is a trivial solution and $\delta W$ does not contain scalar perturbations at all.
This has a major consequence. Just looking at action (\ref{delta2Sinf}) we see that the scalar perturbations during inflation inflation at the leading order are always governed by a local $R^2$ action irrespectively of what the full gravity theory is. Technically this is seen through the simplification of all the system of linearized scalar perturbation equations presented in Appendix~\ref{app:linear}. Effectively it behaves the same way as in the pure dS leaving the analysis and the results of \cite{Koshelev:2016xqb} intact. As a particular important consequence we gain
\begin{equation}
	\Phi+\Psi=0\, .
	\label{phipsi0}
\end{equation}
One can see from the explicit expressions in Appendix~\ref{app:linear} that $\delta W$ depends solely on $\Phi+\Psi$ and as such the contribution which would come from the Weyl tensor piece vanishes.

Having said this we can straightforwardly utilize the results of \cite{Craps:2014wga,Koshelev:2016xqb} to write down the  action for a canonical variable which is
\begin{equation}
\delta^{2}S_{scalar}=\int d^{4}x\sqrt{-\bar{g}}\frac{\lambda}{2\mathcal{F}_{1}\bar{R}}\Upsilon\frac{\mathcal{W}\left(\bar{\Box}\right)}{\mathcal{F}_{R}\left(\bar{\Box}\right)}\left(\bar{\Box}-r_{1}\right)\Upsilon\,.\label{R2}
\end{equation}
Here $\Upsilon$ is the canonical variable in question related to Bardeen potentials as $2\Fc_1\bar R\Psi=\Upsilon$. The operator $\mathcal{W}(\bar\Box)$ is
\begin{equation}
	\Wc(\bar\Box)=3\mathcal{\mathcal{F}}_{R}\left(\bar\Box\right)+\left(\bar{R}+3r_{1}\right)\mathcal{Z}_{1}\left(\bar\Box\right)\, .
	\label{Wbox}
\end{equation}
Comparing operator $\Wc$ with the expression for the spin-0 propagator around dS background found in \cite{Biswas:2016egy} we see that for a consistent theory (around the dS background which is the case during inflation) we should demand
\begin{equation}\Wc(\bar\Box)=e^{2\gamma(\bar\Box)}\label{gammascalar}\end{equation}
for some entire function $\gamma(\bar\Box)$.

One may wonder about the denominator $\Fc_R(\bar\Box)$. Naively one would expect that the whole fraction $\Wc(\bar\Box)/\Fc_R(\bar\Box)$ must be an exponent of some entire function to avoid extra poles in the propagator. It is however not always necessary as the denominator would contribute to poles of the propagator only if it has poles on its own. In a particular case that the denominator itself is an entire function or in a situation that would be propagator poles are beyond the domain of validity of our effective theory (in simple words the excitations are heavier than the effective theory scale, in our case heavier than the scale of inflation) one should not worry about the presence of the denominator in the operator function. The detailed analysis of such a situation is presented in \cite{Koshelev:2017ebj}.

Going further we should proceed with the quantization of perturbations and an evaluation of the two-point function in order to deduce the power spectrum and the corresponding scalar spectral tilt. What is intriguing however, accounting the fact that we have to compute the final quantities at the position of the pole for the canonical variable, i.e. at $\bar\Box=r_1$ and using that $\Zc_1(r_1)=0$ we are going to get answers identical to those in a local $R^2$ theory as $\Wc(r_1)/\Fc_R(r_1)=3$. On the other hand this is not a surprise as in the situation when $\zeta=\delta W=0$ we see that action (\ref{delta2Sinf}) is nothing but a second order variation of an action for a local $R^2$ gravity.

Actual results for the scalar power spectrum $\Pc_\Rc$ for $\Rc=\Psi+H\delta R_{GI}/\dot{\bar R}$ and the corresponding scalar tile $n_s$ can be found in \cite{Craps:2014wga,Koshelev:2016xqb} and are as follows 
\begin{equation}
	\left.\Pc_\Rc\right|_{k=aH}= \frac{H^2}{16\pi^2\epsilon^2}\frac{1}{3\lambda\Fc_1\bar{R}}\,,\quad  n_s\equiv \left.\frac{d\ln \Pc_\Rc}{d\ln k}\right|_{k=aH} \approx 1-\frac{2}{N} \,,\label{Psuni}\end{equation}
where $N$ is the number of $e$-folds.

We thus double argued: first, by using the explicit action for the scalar perturbations and, second, by the rederivation of the action for a canonical variable in the scalar sector that irrespectively of what is the general full gravity theory, an inflation would lead always to the same universal predictions in the scalar sector up to the leading order in the slow-roll correction. One however would get absolutely new corrections coming truly from the non-local operators as long as next to leading orders in the slow-roll approximation or higher, i.e. three- or more, -point correlation functions are considered.

\subsection{Tensor perturbations}

Computation of the tensor perturbations was done in \cite{Koshelev:2016xqb} and already accounts the leading slow-roll approximation. The action for the canonical variable is
\begin{equation}
	\delta^2 S_{tensor}=\int d^4x\sqrt{-\bar g}\frac{\lambda\Fc_1\bar R}4h_{\mu\nu}\left(\bar\Box-\frac{\bar R}6\right)\Pc(\bar\Box)h^{\mu\nu}\, ,
	\label{delta2tensor}
\end{equation}
where $h_{\mu\nu}$ is transverse and traceless and the factor 4 instead of 2 in the denominator is useful as one has to multiply further by 2 to account for two polarizations. The extra operator $\Pc(\Box)$ appears because of the original AID operators and reads as
\begin{equation}
	\Pc(\bar\Box)=1+\frac1{\Fc_1\bar R}\left(\bar\Box-\frac{\bar R}3\right)\Fc_W\left(\bar\Box+\frac{\bar R}3\right)\, .
	\label{Poper}
\end{equation}
Noticec that a constant $\Fc_W(\bar\Box)$ results in the second pole in the spin-2 Lagrangian and this is exactly the Weyl ghost observed by Stelle in \cite{Stelle:1976gc,Stelle:1977ry}. Demanding that no new (and necessarily ghost) spin-2 excitations appear we must have either $\Pc(\bar\Box)=\const>0$ or $\Pc(\bar\Box)=e^{-2\omega(\bar\Box)}$ where $\omega(\bar\Box)$ is an entire function in full analogy with $\gamma(\bar\Box)$ in (\ref{gammascalar}). The first choice results in a non-analytic $\Fc_W(\bar\Box)$. The second choice on the one hand evades a ghost but on the other hand proves that only a truly non-local operator can generate a ghost-free spectrum. As the result presence of a non-constant $\omega(\bar\Box)$ is inevitable. The result for the power spectrum of tensor modes without slow-roll corrections as it was got in \cite{Koshelev:2016xqb} is
\begin{equation}
\left.\mathcal{P}_{\mathcal{T}}\right|_{k=aH}=\frac{H^{2}}{\pi^{2}\lambda\mathcal{F}_{1}\bar{R}}e^{-2\omega\left(\frac{\bar{R}}{6\mathcal{M}^{2}}\right)}
\, .\label{hatH-pwt}
\end{equation}
Note that function $\omega(\bar\Box)$ must be evaluated at the position of the pole which is $\bar R/6$ and also we restore $\Mc$, the scale of non-locality in our model.

Here we advance our study in comparison with \cite{Koshelev:2016xqb} by computing also the tensor tilt and finding it up to the leading order in the slow-roll approximation.
However, to achieve this, the next order slow-roll correction may need to
be added to the tensor power spectrum (\ref{hatH-pwt}), since in the case of the
local $R^2$ model the leading term in $\Pc_{\mathcal{T}}$ does not depend on
$N=-\ln k +\const$ \cite{Starobinsky:1983zz} (c.f. the recent paper \cite{Brooker:2016oqa} in this connection,
too).
A careful computation gives
\begin{equation}
\left.\mathcal{P}_{\mathcal{T}}\right|_{k=aH}=\frac{1}{12\pi^{2}\lambda\mathcal{F}_{1}}\left(1-3\epsilon\right)e^{-2\omega\left(\frac{\bar{R}}{6\mathcal{M}^{2}}\right)}\,.\label{h-corrected-1}
\end{equation}
We use this to obtain 
\begin{equation}
	\begin{aligned}n_{t}\equiv\left.\frac{d\ln\mathcal{P}_{\mathcal{T}}}{d\ln k}\right|_{k=aH}&\approx-\frac{d\ln\mathcal{P}_{\mathcal{T}}}{dN}\left(1+\frac1{2N}\right)\\
		&\approx-\frac3{2N^2}+\left(\frac 8N+\frac6{N^2}\right)\frac{\bar{R}}{6\mathcal{M}^{2}}\omega^{\prime}\left(\frac{\bar{R}}{6\mathcal{M}^{2}}\right)\,,
\end{aligned}
\label{modified-tilt}
\end{equation}
where prime is the derivative with respect to the argument and we have used that $\epsilon=1/(2N)$ with $N$ the number of $e$-folds.
Note that if $\omega^{\prime}\left(\frac{\bar{R}}{6\Mc^2}\right)=0$ we
recover the tensor tilt of the Starobinsky model, i.e. $n_{t}=-\frac{3}{2N^{2}}$ which is a red tilt $n_{t}<0$.

\subsection{$r$ and modified consistency relation}

Using a standard (local) results for the scalar power spectrum as advocated above (\ref{Psuni}) and modified tensor power spectrum (\ref{hatH-pwt}) the tensor to scalar ratio is given by 
\begin{equation}
	r=\left.\frac{\mathcal{P}_{\mathcal{T}}}{\mathcal{P}_{\mathcal{R}}}\right|_{k=aH}=\frac{12}{N^2}e^{-2\omega\left(\frac{\bar{R}}{6\mathcal{M}^{2}}\right)}\, .\label{T2Sratio}
\end{equation}

Therefore the presence of the Weyl tensor squared term in the action modifies
the single field consistency relation $\left(r=-8n_{t}\right)$
as follows
\begin{equation}
	\frac{r}{n_{t}}=-8\frac{e^{-2\omega\left(\frac{\bar{R}}{6\mathcal{M}^{2}}\right)}}{1-\frac{16}3N\frac{\bar R}{6\mathcal{M}^{2}}\omega^{\prime}\left(\frac{\bar{R}}{6\mathcal{M}^{2}}\right)}\, .\label{modified-tilt-1}
\end{equation}

Using the fact that our computations do not depend technically on whether operators are of finite or infinite order in derivatives one can readily compare our results with an analogous derivation done in \cite{Baumann:2015xxa} where a pure Weyl tensor squared term was considered. In that case our answers can be shown to match. We however stress again that only a truly non-local operator, i.e. AID operator is necessary to get rid of the Weyl ghost.

\section{Conclusions}
\label{Concl}

$R^2$ always stood as the most successful theory of inflation and it is now the best fit for the most recent Planck data. In a recent study \cite{Koshelev:2016xqb} this model realized in the context of non-local gravity that was shown to be UV complete in the sense of having no ghosts (unitarity) and being super-renormalizable (or finite). This was the significant theoretical development which embeds $R^2$ inflationary paradigm in a finite theory of quantum gravity. The present paper further extends this previous study with more rigorous analysis of the action, generalized solutions for EOM and derivation of inflationary parameters that can be tested in the future CMB data. 
 
We have started by deriving explicitly that a most general theory of gravity that contributes to the linearly perturbed EOM around MSS contains the Einstein-Hilbert term, $R^2$ and Weyl tensor squared terms with AID operators in between, and the cosmological constant, see (\ref{actionour4}) and (\ref{actionourD}). Using the power of Bianchi identities we were able to reduce the final action presented in \cite{Biswas:2016egy} effectively eliminating the Ricci tensor squared term. 
Our proof applies to any theory of gravity in any dimension. It is worth mentioning that SFT \cite{Arefeva:2001ps} provides a natural motivation for such kind of AID actions. 

We have proceeded by presenting a rigorous mathematical proof that the trace equation of a local $R^2$ gravity without the cosmological term i.e., $\Box R=r_1 R$ is the only solution to EOM of our AID quadratic gravity also without the cosmological term as long as spatially flat FLRW metrics are considered.
This means that even though we have complicated the local higher derivative theory of gravity \cite{Stelle:1976gc} with AID operators, the space of background solutions remains the same satisfying $\Box R=r_1 R$. In the situation when the Weyl tensor term is not included from the very beginning the claim remains true for any metric which is space-homogeneous in the synchronous reference frame, for example, anisotropic metrics, in particular Bianchi~I type configurations, etc.

Further, we have derived the full perturbed second order action of AID quadratic gravity without the cosmological term around general backgrounds satisfying $\Box R=r_1 R$ and conditions (\ref{conditions}), significantly boosting the previous studies where perturbations were only computed around MSS \cite{Biswas:2012bp,Craps:2014wga,Koshelev:2016xqb}. This full second order action is undoubtedly useful for further studies in the framework of AID gravities.

One of the crucial immediate task to be done in a forthcoming study is to extend these achievements to the models involving the cosmological term and to other types of metric, for example space-inhomogeneous or those describing spherically symmetric solution.
This will allow one to attack in full, for example, the study of non-singular and ghost-free bounce configurations \cite{Biswas:2012bp,Koshelev:2013lfm} which require the cosmological term to be present in the action. Also one can use the AID quadratic gravity framework to reconsider the problem of the curvature singularity which was proven to be generic in the case of Bianchi~I metric in a local $R^2$ model \cite{Muller:2017nxg}.

Using the above described tools which are constructed in the present paper for the first time, we have come with the inflationary predictions of AID quadratic gravity model such as scalar spectral index, tensor to scalar ratio and tensor tilt consistently computing them in the leading order of the slow-roll approximation.

On this way we have proven that in the leading slow-roll approximation scalar perturbations in our model are equivalent to those in a local $R^2$ gravity. Our analysis thus makes it transparent that the scalar power spectrum remains the same as in a local $R^2$ model proving therefore that our model is to be the best fit with the present constraint $n_s=0.968\pm 0.006$. 

The tensor power spectrum however gets modified exactly due to a differential operator in the Weyl tensor squared term introducing thereby a new parameter associated solely with this operator. As a result, the tensor to scalar ratio $r$ gets a correction by a parameter that can give any value of $r<0.07$ following (\ref{T2Sratio}). As an interesting but not a surprising consequence the computed tensor tilt deviates from a local $R^2$ model and thus the consistency relation gets modified as in (\ref{modified-tilt-1}).
This resembles the results obtained in \cite{Baumann:2015xxa} with a huge difference that our model can avoid ghosts by promoting an operator in the Weyl tensor squared term to an AID operator. In particular, Eq. (\ref{Poper}) and the discussion thereafter gives a very simple and clear explanation that the only way to defeat the Weyl ghost is to introduce a truly infinite derivative operator as long as one allows only analytic dependence on derivatives.

Our current AID quadratic gravity model modifies the tensor power spectrum and consequently $r$ by a new parameter which is associated in this model with a quantum gravity prescription in the UV regime. This is in contrast to many other ``Starobinsky''-like models in the market \cite{Ellis:2013nxa,Kehagias:2013mya,Koshelev:2016vhi,Burgess:2016owb} which modify only the scalar power spectrum. 
The tensor tilt in our model gets a new parameter related to the scale of non-locality $\Mc$. The value of this new parameter can be fixed by future observations of primordial $B$-modes. Therefore, inflation in AID quadratic gravity meets all the current CMB constraints by PLANCK and is undoubtedly a very interesting and natural target for future CMB probes. 
We emphasize also that despite the fact that we can have any value of $r<0.07$ in this model the energy scale of inflation remains the same as the latter is determined by scalar perturbations. This is a noteworthy feature of our model which is absent in the scalar field attractor models, i.e. so called $\alpha$-attractor models \cite{Kallosh:2013yoa}.

Our results for AID quadratic gravity theory provide a foundation for studying not only inflation but also bounce, black holes, late time acceleration etc. in this framework. Given the theoretical progress we have achieved in the present paper future studying of reheating, non-gaussianities and other crucial questions are very important and timely. Intensifying the study of more inflationary parameters in combination with constraints from the observational camp would allow to narrow, for instance, the scale of non-locality and to start shaping the non-perturbative quantum gravity.

\acknowledgments

We thank Jo\~ao Marto, Anupam Mazumdar and Leonardo Modesto for useful discussions.
AK is supported by FCT Portugal investigator project IF/01607/2015 and FCT
Portugal fellowship  SFRH/BPD/105212/2014.
KSK acknowledges the support from the FCT PhD grant 
SFRH/BD/51980/2012. AK and KSK are supported in part by FCT Portugal grant UID/MAT/00212/2013
and COST Action CA15117 (CANTATA). AS is supported by the RSF grant 16-12-10401.

\appendix

\section{Notations and common quantities}

\label{ap:notation} 


The metric $g_{\mu\nu}$ is such that
\begin{equation}
	g_{\mu\nu}=(-,+,+,+,\dots),\quad g_{\mu\nu}g^{\mu\nu}=D\,.
\end{equation}
In most cases $D=4$ unless indicated otherwise.
Small Greek letters are the $D$-dimensional indices.
\begin{eqnarray}
\Gamma_{\mu\nu}^{\rho}=\frac{1}{2}g^{\rho\sigma}(\pd_{\mu}g_{\nu\sigma}+\pd_{\nu}g_{\mu\sigma}-\pd_{\sigma}g_{\mu\nu})\,,~\cpd_{\mu}F_{.\beta.}^{.\alpha.}=\pd_{\mu}F_{.\beta.}^{.\alpha.}+\Gamma_{\mu\chi}^{\alpha}F_{.\beta.}^{.\chi.}-\Gamma_{\mu\beta}^{\chi}F_{.\chi.}^{.\alpha.}\,.
\end{eqnarray}
\begin{equation}
R_{\mu\nu\rho}^{\sigma} = \pd_{\nu}\Gamma_{\mu\rho}^{\sigma}-\pd_{\rho}\Gamma_{\mu\nu}^{\sigma}+\Gamma_{\chi\nu}^{\sigma}\Gamma_{\mu\rho}^{\chi}-\Gamma_{\chi\rho}^{\sigma}\Gamma_{\mu\nu}^{\chi}\,.
\end{equation}
\begin{equation}
\begin{split}R_{\mu\nu\rho\sigma}=-R_{\mu\nu\sigma\rho}=-R_{\nu\mu\rho\sigma} & =R_{\rho\sigma\mu\nu}\,,\quad R_{\mu\nu\rho\sigma}+R_{\mu\sigma\nu\rho}+R_{\mu\rho\sigma\nu}=0\,.\end{split}
\label{sym:riem}
\end{equation}
The last property in the latter line is sometimes called algebraic
Bianchi identity.
\begin{equation}
R_{\mu\nu}=R_{\mu\sigma\nu}^{\sigma}\,,\quad R_{\mu\nu}=R_{\nu\mu}\,,\quad R=R_{\mu}^{\mu}\label{def:ricci}
\end{equation}
\begin{eqnarray}
[\cpd_{\mu},\cpd_{\nu}]A_{\rho}=R_{\rho\nu\mu}^{\chi}A_{\chi}\,,\quad\Box=g^{\mu\nu}\cpd_{\mu}\cpd_{\nu}\,.
\end{eqnarray}
The (differential) Bianchi identity is: 
\begin{equation}
\cpd_{\lambda}R_{\mu\nu\rho\sigma}+\cpd_{\sigma}R_{\mu\nu\lambda\rho}+\cpd_{\rho}R_{\mu\nu\sigma\lambda}\equiv0\,.\label{bianchi}
\end{equation}

We note down the following important second rank tensors 
\begin{eqnarray}
	\text{Einstein: }G_{\mu\nu} & = & R_{\mu\nu}-\frac{1}{2}Rg_{\mu\nu}\,,\text{ due to Bianchi identity }\nabla_{\mu}G_{\nu}^{\mu}\equiv0\,,\label{bianchi:G}\\
\text{Schouten: }S_{\mu\nu} & = & \frac{1}{D-2}\left(R_{\mu\nu}-\frac{1}{2(D-1)}Rg_{\mu\nu}\right)\,,\\
\text{Traceless Ricci: }L_{\mu\nu} & = & R_{\mu\nu}-\frac{1}{D}g_{\mu\nu}R\,,\quad L_{\mu}^{\mu}=0\,.
\end{eqnarray}
All these tensors are symmetric.

An important third rank tensor is the Cotton tensor: 
\begin{equation}
C_{\mu\nu\alpha} = \nabla_{\mu}S_{\nu\alpha}-\nabla_{\nu}S_{\mu\alpha},~C_{\mu\nu\alpha}+C_{\alpha\mu\nu}+C_{\nu\alpha\mu}=0,~ C_{\mu\alpha}^{\phantom{\mu\alpha}\alpha}=0,~\nabla^{\alpha}C_{\mu\nu\alpha}=0
\end{equation}

The fourth rank Weyl tensor is: 
\begin{eqnarray}
W_{\alpha\nu\beta}^{\mu} & = & R_{\alpha\nu\beta}^{\mu}-\frac{1}{D-2}(\delta_{\nu}^{\mu}R_{\alpha\beta}-\delta_{\beta}^{\mu}R_{\alpha\nu}+{g}_{\alpha\beta}R_{\nu}^{\mu}-{g}_{\alpha\nu}R_{\beta}^{\mu})\nonumber \\
 & + & \frac{R}{(D-2)(D-1)}(\delta_{\nu}^{\mu}{g}_{\alpha\beta}-\delta_{\beta}^{\mu}{g}_{\alpha\nu})=\nonumber\\
& = & R_{\alpha\nu\beta}^{\mu}-\delta_{\nu}^{\mu}S_{\alpha\beta}+\delta_{\beta}^{\mu}S_{\alpha\nu}-g_{\alpha\beta}S_{\nu}^{\mu}+g_{\alpha\nu}S_{\beta}^{\mu}\,.
\end{eqnarray}
The Weyl tensor has all the symmetry properties of the Riemann tensor
(\ref{sym:riem}) and it is absolutely traceless, i.e. 
$
W_{\alpha\mu\beta}^{\mu}=0\,.
$
Moreover it is invariant under the conformal rescaling, i.e. 
\begin{eqnarray}
\hat{W}_{\alpha\beta\gamma}^{\mu}=W_{\alpha\beta\gamma}^{\mu}\:\text{ for }\:\hat{g}_{\mu\nu}=\Omega^{2}(x)g_{\mu\nu}\,.
\end{eqnarray}
This implies that the Weyl tensor is zero on conformally flat
manifolds (i.e. when the metric can have the form $ds^{2}=a(x)^{2}\eta_{\mu\nu}dx^{\mu}dx^{\nu}$ where $\eta_{\mu\nu}$
is the Minkowski metric with the same signature).

In fact, one should keep in mind that in $D\geq4$ vanishing Weyl tensor is a necessary and a sufficient condition for the space-time to be conformally flat.

Applying the Bianchi identity to the Weyl tensor one can find 
\begin{equation}
\nabla^{\beta}W_{\alpha\mu\nu\beta}=-({D-3})C_{\alpha\mu\nu}=-(D-3)(\nabla_{\alpha}S_{\mu\nu}-\nabla_{\mu}S_{\alpha\nu})\,.
\end{equation}
Squaring the last equality we get 
\begin{equation}
(\nabla^{\beta}W_{\alpha\mu\nu\beta})^{2}=-2({D-3})\nabla^{\beta}W_{\alpha\mu\nu\beta}\nabla_{\alpha}S_{\mu\nu}\,.\label{bianchidw2}
\end{equation}
Next one can compute 
\begin{equation}
\begin{split}\frac{1}{D-3}\nabla^{\alpha}\nabla^{\beta}W_{\alpha\mu\nu\beta} & =-\Box S_{\mu\nu}+\frac{1}{2(D-1)}\nabla_{\mu}\pd_{\nu}R+\frac{1}{D-2}W_{\rho\nu\mu\alpha}L^{\alpha\rho}\\
 & +\frac{D}{(D-2)^{2}}L_{\mu\alpha}L_{\nu}^{\alpha}-\frac{1}{(D-2)^{2}}g_{\mu\nu}L_{\alpha\beta}^{2}+\frac{1}{(D-1)(D-2)}RL_{\mu\nu}
\end{split}
\label{bianchid2w}
\end{equation}
Going further one finds 
\begin{equation}
\begin{split}\nabla^{\mu}\nabla^{\alpha}\nabla^{\beta}W_{\alpha\mu\nu\beta} & =(D-3)\nabla^{\mu}(W_{\mu\nu\rho}^{\alpha}S_{\alpha}^{\rho})+(D-4)\nabla_{\alpha}W_{\nu\mu\alpha}^{\rho}S_{\rho}^{\mu}\end{split}
\end{equation}
We see that in 4 dimensions the combination 
\begin{equation}
\begin{split}B_{\mu\nu}=\nabla^{\alpha}\nabla^{\beta}W_{\alpha\mu\nu\beta}-\frac{1}{2}W_{\mu\nu\rho}^{\alpha}R_{\alpha}^{\rho}\end{split}
\label{def:bach}
\end{equation}
is transverse. We have used here the normalization of the Schouten
tensor and the traceless property of the Weyl tensor. The latter combination
is named Bach tensor. It is symmetric, traceless, transverse (for
$D=4$) and has the conformal weight $-2$ (i.e. scales as $\Omega(x)^{-2}$ upon scaling
of the metric by $\Omega(x)^2$).
	
The spatially flat FLRW Universe metric is 
\begin{eqnarray}
ds^{2}=-dt^{2}+a(t)^{2}\left({dr^{2}}+r^{2}d\Omega^{2}\right)\,.\label{frwcosmic}
\end{eqnarray}
$t$ is the cosmic time and $a(t)$ is the scale factor. We intrinsically assume zero spatial curvature.
The Hubble function is $H=\dot{a}/a$ with dot denoting
the derivative with respect to $t$.
Equivalently we write the FLRW metric (\ref{frwcosmic}) as 
\begin{eqnarray}
ds^{2}=a(\tau)^{2}\left(-d\tau^{2}+\delta_{ij}dx^{i}dx^{j}\right)\,.\label{frwtau}
\end{eqnarray}
$\tau$ is the conformal time such that $ad\tau=dt$. Spatially flat FLRW Universe is conformally
flat and the Weyl tensor in it is identically zero. The background
quantities in the latter metric are 
\begin{eqnarray}
\Gamma_{0\nu}^{\mu} & = & \Hc\delta_{\nu}^{\mu}\,,\quad\Gamma_{\mu\nu}^{0}=\delta_{\mu\nu}\Hc\,,\quad\Hc=a'/a\,,\\
R & = & \frac{6}{a^{2}}(\Hc'+\Hc^{2})\,,\quad R_{\mu\nu}=\left(\begin{array}{cc}
-3\Hc' & 0\\
0 & (\Hc'+2\Hc^{2})\delta_{ij}
\end{array}\right)\,,\\
R_{i0j}^{0} & = & \Hc'\delta_{ij}\,,\quad R_{0j0}^{i}=-\Hc'\delta_{j}^{i}\,,\quad R_{jkm}^{i}=\Hc^{2}(\delta_{k}^{i}\delta_{jm}-\delta_{m}^{i}\delta_{kj})\,.\label{frwtaucscurves}
\end{eqnarray}
We use the index ``$0$'' for the $\tau$-component of any tensor in order not to confuse with just a small Greek letter (the
cosmic time is used less often and wherever needed we designate it
with the index $t$). Latin small letters from the middle of the alphabet
are spatial indices, prime is the derivative
with respect to the conformal time $\tau$. Further: 
\begin{eqnarray}
\pd_{0} & = & a\pd_{t},~ H=\Hc/a,~\Box_{s}  =  -\frac{1}{a^{2}}(\pd_{0}^{2}+2\Hc\pd_{0}-\delta^{ij}\pd_{i}\pd_{j}),~
R = 6\dot{H}+12H^{2}\,, 
\end{eqnarray}
where $\Box_{s}$ is the d'Alembertian operator acting on scalars.

In using non-local operators $\Fc_X(\Box)$ where $X$ is some notational index we always assume that these operators are analytic functions of their arguments. This allows to write them in a Taylor series representation
\begin{equation}
	\Fc_X(\Box)=\sum_{n=0}f_{Xn}\Box^n
	\label{Fseries}
\end{equation}


The metric perturbations are introduced as
\begin{equation}
	g_{\mu\nu}=\bar g_{\mu\nu}+h_{\mu\nu},\quad h=h^\mu_\mu
	\label{hmunu}
\end{equation}
For any other quantity $\varphi$ apart from the metric we use $\bar\varphi$ for its background value and $\delta\varphi$ for linear corrections.
A spatial Fourier transform used to study perturbations is
\begin{equation}
\varphi(\tau,\vec{x})=\int\varphi(\tau,\vec{k})e^{i\vec{k}\vec{x}}d\vec{k},~
\bar{\Box}_{k}\to-\frac{1}{a^{2}}(\pd_{0}^{2}+2\Hc\pd_{0}+k^{2})\,.\label{frwboxsk}
\end{equation}
$\vec k$ is the spatial comoving momentum.

\section{Reduction of (\ref{actionverygeneral2LC}) to (\ref{actionour4})}
\label{app:reduction}

As the first step of this procedure we notice that one can use Schouten tensor $S_{\mu\nu}$
instead of $L_{\mu\nu}$ in (\ref{actionverygeneral2LC}). This is a linear transformation of $L$-tensor
and it leads to some redefinition of function $\tilde{\Fc}_{R}(\Box)$.
Using that $\Fc_{L}(\Box)$ is an analytic function we can write it
as a Taylor series and the corresponding expression is 
\begin{equation}
\intdg{4}S_{\mu\nu}\Fc_{L}(\Box)S^{\mu\nu}=\intdg{4}S_{\mu\nu}\sum_{n\geq0}f_{Ln}\Box^{n}S^{\mu\nu}\,.\label{boxStrickrecurs}
\end{equation}
The zero term in this series can be dropped thanks to the presence
of the Gauss-Bonnet (GB) invariant in 4 dimensions. With $\Box$-s
in between there is no such an obvious possibility. However, we can
write the series without the zero term as follows 
\[
\intdg{4}(\Box S_{\mu\nu})\sum_{n\geq1}f_{Ln}\Box^{(n-1)}S^{\mu\nu}\,.
\]
We have moved one d'Alembertian to the left using an integration by
parts since we are doing our computation under the integral.

The second step is to express $\Box S_{\mu\nu}$ using Bianchi identity
(\ref{bianchid2w}) and the right most $S_{\mu\nu}$ through $L_{\mu\nu}$
and $g_{\mu\nu}R$. Schematically without explicit coefficients this
can be written as 
\begin{equation}
\begin{split}\intdg{4} & \left(x_{1}\nabla^{\alpha}\nabla^{\beta}W_{\alpha\mu\phantom{\nu}\beta}^{\phantom{\alpha\mu}\nu}+x_{2}\nabla_{\mu}\pd^{\nu}R+x_{3}W_{\alpha\mu\phantom{\nu}\beta}^{\phantom{\alpha\mu}\nu}L^{\alpha\beta}+x_{4}L_{\mu\alpha}L^{\alpha\nu}+x_{5}\delta_{\mu}^{\nu}L_{\alpha\beta}^{2}\right.\\
 & \left.+x_{6}L_{\mu}^{\nu}R\right)\sum_{n\geq1}f_{Ln}\Box^{(n-1)}\left(y_{1}L_{\nu}^{\mu}+y_{2}\delta_{\nu}^{\mu}R\right)\,.
\end{split}
\label{boxStrick}
\end{equation}
Here $x_{i},~y_{i}$ are numeric coefficients. We observe the following
by primary inspection: terms proportional to $\sim WLL$ or $\sim LLL$
would not survive under the second variation around MSS as they contain
3 multipliers which are trivial on the background; since Weyl tensor
and $L$-tensor are fully traceless any contraction of them with $\delta$-symbol
vanishes. Terms which still can contribute are written below: 
\begin{equation}
\begin{split}\intdg{4} & \left[\left(x_{1}\nabla^{\alpha}\nabla^{\beta}W_{\alpha\mu\phantom{\nu}\beta}^{\phantom{\alpha\mu}\nu}+x_{2}\nabla_{\mu}\pd^{\nu}R+x_{6}L_{\mu}^{\nu}R\right)\sum_{n\geq1}f_{Ln}\Box^{(n-1)}y_{1}L_{\nu}^{\mu}\right.\\
 & +\left.\left(x_{2}\Box R+(x_{4}+4x_{5})L_{\alpha\beta}^{2}\right)\sum_{n\geq1}f_{Ln}\Box^{(n-1)}y_{2}R\right]\,.
\end{split}
\label{boxStrick2}
\end{equation}

A careful examination reveals that the last term in the first line
of the latter expression can survive the second variation only if
each of two $L$-tensors is varied. As such this term with $R$ taking its
background value $\bar{R}$ resembles the $L$-piece in action (\ref{actionverygeneral2LC}).
We thus can tackle this term recursively looping back to (\ref{boxStrickrecurs})
and considering a new expression 
\begin{equation}
\intdg{4}x_{6}y_{1}\bar{R}S_{\mu\nu}\sum_{n\geq0}f_{L(n+1)}\Box^{n}S^{\mu\nu}\,.\label{boxStrickrecursnext}
\end{equation}
Obviously, this is essentially the same expression as in (\ref{boxStrickrecurs})
with new series coefficients. Iterating more and more (infinitely
many) times we eliminate this term entirely. At each iteration the
zero term with $n=0$ can be dropped thanks to the use of the GB invariant.

Further, first term on the second line in (\ref{boxStrick2}) can
be effectively absorbed in $R\tilde{\Fc}_{R}(\Box)R$ piece. The second
term of the last line only contributes to a second variation when
both $L$-tensor multipliers are varied and $R$ takes its background value.
Thus, it is again a local $L^{2}$ term and we drop it thanks to the
use of the GB invariant.

So for the moment what we are left with is just first two terms from
the first line of (\ref{boxStrick2}). We rewrite these two terms
for convenience: 
\begin{equation}
\begin{split}\intdg{4} & \left(x_{1}\nabla^{\alpha}\nabla^{\beta}W_{\alpha\mu\phantom{\nu}\beta}^{\phantom{\alpha\mu}\nu}+x_{2}\nabla_{\mu}\pd^{\nu}R\right)\sum_{n\geq1}f_{Ln}\Box^{(n-1)}y_{1}L_{\nu}^{\mu}\,.\end{split}
\label{boxStrick3}
\end{equation}
The second term here is somewhat simpler to play with and we will show
using its example that one can commute covariant derivatives without
paying attention to extra contribution exactly like it happens in
Minkowski background. For the zero term in the series we simply write
(modulo the constant coefficient $x_{2}f_{L1}y_{1}$) 
\begin{equation}
\begin{split}\intdg{4} & (\nabla_{\mu}\pd^{\nu}R)L_{\nu}^{\mu}=-\intdg{4}(\pd^{\nu}R)\nabla_{\mu}(G_{\nu}^{\mu}+\frac{1}{4}\delta_{\nu}^{\mu}R)=\frac{1}{4}\intdg{4}R\Box R\,,\end{split}
\label{boxStrick4}
\end{equation}
where we have used the Bianchi identity for the Einstein tensor (\ref{bianchi:G}). So
this is just another contribution to $R\tilde{\Fc}_{R}(\Box)R$ piece.
Given we have one d'Alembertian operator we can write the following
(again we omit the constant coefficient $x_{2}f_{L2}y_{1}$) 
\begin{equation}
\begin{split}\intdg{4} & (\nabla_{\mu}\pd^{\nu}R)\Box L_{\nu}^{\mu}=-\intdg{4}(\pd^{\nu}R)\nabla_{\mu}\Box L_{\nu}^{\mu}\\
= & \intdg{4}(\pd^{\nu}R)\left(\Box\nabla_{\mu}L_{\nu}^{\mu}+\nabla^{\alpha}R_{\rho\alpha}L_{\nu}^{\rho}+\nabla^{\alpha}R_{\nu\alpha\mu}^{\rho}L_{\rho}^{\mu}+R_{\nu\alpha\mu}^{\rho}\nabla^{\alpha}L_{\rho}^{\mu}\right)\,.
\end{split}
\label{boxStrick5}
\end{equation}
The first term on the right can be treated in full analogy with (\ref{boxStrick4})
with one extra d'Alembertian inside. All other pieces can be in fact
can be treated again as (\ref{boxStrick4}). Indeed, a non-zero second
variation would only be present if $R$ in the first factor $\pd^{\nu}R$
and $L$-tensor in the second factor are varied. As such, Riemann
and Ricci tensors take their background values and all the terms but
first in the second parenthesis can be absorbed in the zero series
term. With the same strategy in mind, for any $n$ the $n$-th term
in the series can be redistributed into $(n-1)$-th term and $R\tilde{\Fc}_{R}(\Box)R$.
Finally this recursion absorbs everything of interest into $R\tilde{\Fc}_{R}(\Box)R$.

The first term in (\ref{boxStrick3}) in fact produces indeed non-vanishing
remaining contributions. For the first term in series expansion without
d'Alembertian we have (we again omit constant coefficient $x_{1}f_{L1}y_{1}$
below) 
\begin{equation}
\begin{split}\intdg{4} & (\nabla^{\alpha}\nabla^{\beta}W_{\alpha\mu\phantom{\nu}\beta}^{\phantom{\alpha\mu}\nu})L_{\nu}^{\mu}\,.\end{split}
\label{boxStrick6}
\end{equation}
Given we have at least one d'Alembertian, we will first replace $L_{\mu\nu}\to S_{\mu\nu}$
which is possible since the difference is proportional to the metric
and the Weyl tensor is traceless, and second, we will use (\ref{bianchid2w})
to express $\Box S_{\mu\nu}$ like in (\ref{boxStrick}). This yields
(modulo overall constant multiplier $x_{1}y_{1}$) 
\begin{equation}
	\begin{split}\intdg{4} & (\nabla^{\alpha}\nabla^{\beta}W_{\alpha\mu\phantom{\nu}\beta}^{\phantom{\alpha\mu}\nu})\sum_{m\geq0}f_{L(m+2)}\Box^{m}\Box L^{\mu}_{\nu}\\
		=\intdg{4} & (\nabla^{\alpha}\nabla^{\beta}W_{\alpha\mu\phantom{\nu}\beta}^{\phantom{\alpha\mu}\nu})\sum_{m\geq0}f_{L(m+2)}\Box^{m}\left(x_{1}\nabla^{\rho}\nabla^{\sigma}W_{\rho\phantom{\mu}\nu\sigma}^{\phantom{\rho}\mu}+x_{2}\nabla^{\mu}\pd_{\nu}R\right.\\
		& +x_{3}W_{\rho\phantom{\mu}\nu\sigma}^{\phantom{\rho}\mu}L^{\rho\sigma}+x_{4}L_{\nu\rho}L^{\rho\mu}+x_{5}\delta_{\nu}^{\mu}L_{\rho\sigma}^{2}\left.+x_{6}L_{\nu}^{\mu}R\right)\,.
\end{split}
\label{boxStrick7}
\end{equation}
Terms $\sim WLL$ or $\sim LLL$ cannot contribute to the quadratic
variation as they are cubic in quantities which are identically zero
on the background. The term $x_{6}L_{\nu}^{\mu}R$ can only survive having $R$
taking its background value $\bar{R}$. Otherwise, the second variation
of this term in the action will be zero. As such, this term resembles
the LHS of the equality with one d'Alembertian less. One can recursevely
apply (\ref{bianchid2w}) as long as no d'Alembertians are left at
all. Without d'Alembertian oeprators in between the structure like
in (\ref{boxStrick6}) is reproduced.

The term $x_{2}\nabla_{\nu}\pd^{\mu}R$ can be transformed as follows
\begin{equation}
\begin{split}\intdg{4} & (\nabla^{\alpha}\nabla^{\beta}W_{\alpha\mu\phantom{\nu}\beta}^{\phantom{\alpha\mu}\nu})\sum_{m\geq0}f_{L(m+2)}\Box^{m}x_{2}\nabla^{\mu}\pd_{\nu}R\\
\to-\intdg{4} & (\nabla^{\mu}\nabla^{\alpha}\nabla^{\beta}W_{\alpha\mu\phantom{\nu}\beta}^{\phantom{\alpha\mu}\nu})\sum_{m\geq0}f_{L(m+2)}\Box^{m}x_{2}\pd_{\nu}R\\
=\intdg{4} & (\nabla^{\mu}(B_{\mu}^{\nu}-\frac{1}{2}W_{\alpha\mu\phantom{\nu}\beta}^{\phantom{\alpha\mu}\nu}L^{\alpha\beta}))\sum_{m\geq0}f_{L(m+2)}\Box^{m}x_{2}\pd_{\nu}R\,.
\end{split}
\label{boxStrick8}
\end{equation}
The first transformation is possible thanks to the same logic as explained
after (\ref{boxStrick5}). According to it we move derivatives absorbing
all newly emerging Riemann tensor terms inside of terms with one d'Alembertian
operator less. The second transform is an equality which uses the
definition of the Bach tensor $B_{\mu\nu}$ (see (\ref{def:bach})). This tensor is transverse and as such this contribution
vanish. This is exactly like that in $D=4$. In higher dimensions the divergence of the analog of Bach tensor gains non-vanishing contributions of the form $\sim WL$ and as such similar in their properties to the second term which is already present. This second term (and possible its complements in $D>4$) produces $\sim WL\pd R$ contribution which does
not survive the second variation as all three factors must be varied to be non-trivial.

The last unaccounted term in (\ref{boxStrick7}) which actually survives
is 
\begin{equation}
	\begin{split}\intdg{4} & (\nabla^{\alpha}\nabla^{\beta}W_{\alpha\mu\phantom{\nu}\beta}^{\phantom{\alpha\mu}\nu})\sum_{m\geq0}f_{L(m+2)}\Box^{m}x_{1}\nabla^{\rho}\nabla^{\sigma}W_{\rho\phantom{\mu}\nu\sigma}^{\phantom{\rho}\mu}\,.\end{split}
\label{boxStrick9}
\end{equation}

Thus to the moment we have the following result. We started with (\ref{boxStrickrecurs})
and ended up with two non-vanishing contributions (\ref{boxStrick6})
and (\ref{boxStrick9}) while also function $\tilde{\Fc}_{R}(\Box)$
got redefined due to absorption of similar terms into it.

In order to simplify (\ref{boxStrick6}) we move one derivative by
integration by parts to $L$-tensor and use (\ref{bianchidw2}) to
write an equivalent expression (modulo a constant factor): 
\begin{equation}
\intdg{4}(\nabla^{\beta}W_{\alpha\mu\nu\beta})\nabla_{\gamma}W^{\alpha\mu\nu\gamma}=\intdg{4}(\nabla_{\gamma}W_{\alpha\mu\nu\beta})\nabla^{\beta}R^{\alpha\mu\nu\gamma}\,.\label{boxStrick10}
\end{equation}
Here we followed the logic explained after (\ref{boxStrick5}) according
to which we can move derivatives like in Minkowski space. We also
have replaced one Weyl tensor with the Riemann tensor as their difference
vanishes upon contraction with another (traceless) Weyl tensor. We
can now apply the Bianchi identity (\ref{bianchi}) and transform
the last expression to 
\begin{equation}
\begin{split} & -\intdg{4}(\nabla_{\gamma}W_{\alpha\mu\nu\beta})(\nabla^{\gamma}R^{\alpha\mu\beta\nu}+\nabla^{\nu}R^{\alpha\mu\gamma\beta})\\
= & \intdg{4}(-W_{\alpha\mu\nu\beta}\Box W^{\alpha\mu\nu\beta}-\nabla^{\nu}W_{\alpha\mu\nu\beta}\nabla_{\gamma}W^{\alpha\mu\gamma\beta})\,.
\end{split}
\label{boxStrick11}
\end{equation}
Here we moved derivatives like in Minkowski space again and returned
to the Weyl tensor as this transformation is identical. We see that
the last term in the last expression is just the LHS in (\ref{boxStrick10})
with an opposite sign. Therefore, expression in (\ref{boxStrick6})
is equivalent modulo a constant factor to 
\begin{equation}
\intdg{4}W_{\alpha\mu\nu\beta}\Box W^{\alpha\mu\nu\beta}\,,\label{bosStrick12}
\end{equation}
and can be absorbed inside of $\tilde{\Fc}_{W}(\Box)$.

In order to simplify (\ref{boxStrick9}) we have to repeat the same
sequence of moves like for (\ref{boxStrick6}) just four times longer.
The bottom line is that 
\begin{equation}
	\begin{split}\intdg{4} & (\nabla^{\alpha}\nabla^{\beta}W_{\alpha\mu\phantom{\nu}\beta}^{\phantom{\alpha\mu}\nu})\Box^{m}\nabla^{\rho}\nabla^{\sigma}W_{\rho\phantom{\mu}\nu\sigma}^{\phantom{\rho}\mu}\to\intdg{4}W_{\alpha\mu{\nu}\beta}\Box^{m+2}W^{\alpha{\mu}\nu\beta}\,,\end{split}
\label{boxStrick13}
\end{equation}
again effectively redefining $\tilde{\Fc}_{W}(\Box)$.

Summing all up we come to (\ref{actionour4}).

\section{Solving (\ref{tEOMtrace1}) without using (\ref{conditions})}
\label{wasansdiff:app}
Let us find all solutions of Eq. (\ref{tEOMtrace1})
\beq 
AR=B(\dot R^2+2r_1R^2),~~A=M_P^2-6\la r_1{\cF}_R(r_1)\, ,
~~B=\la{\cF}_R^{(1)}(r_1)
\label{trace}
\eeq
under the condition that $R$ satisfies the equation (\ref{wasans})
\beq
\B R = r_1R
\label{trace1}
\eeq
with $r_1>0$ and a spatially flat FLRW space-time is assumed. The simplest way to solve this problem is to reduce the differential equation (\ref{trace}), which is the second order with respect to the Hubble function 
$H\equiv \dot a/a$ (the scale factor $a(t)$ itself does not enter due to invariance under rescaling of all spatial coordinates),  to an expression containing $R(t)$ only which should 
be an {\em identity} if $R$ is not a constant.

Note first that one such solution is $R\equiv 0$, and then no conditions on $A,B$ and $r_1$ arise.

Let us assume further that $R$ is not identically equal to $0$. Let $B\not= 0$. Then
\beq
\dot R=-\sqrt{R\left(\frac{A}{B}-2r_1R\right)}\, ,
\label{dotR1}
\eeq
where we take the minus sign for $\dot R$ corresponding to decrease of $R$ with time for definiteness.  If $r_1\not= 0$ and $R\not= 0$, $\dot R\not= 0$ too. By differentiating Eq.~(\ref{trace}) with respect to time and dividing both its sides by $\dot R$, one gets
\beq
\frac{A}{2B}=\ddot R + 2r_1R=-3H\dot R+r_1R~.
\label{dotR2}
\eeq
From Eqs. (\ref{dotR1}) and (\ref{dotR2}), the expression for $H(t)$ follows:
\beq
H=\frac{1}{6}\sqrt{\frac{A}{BR}-2r_1}~.  
\label{H}
\eeq
Differentiating Eq. (\ref{H}) and using Eq. (\ref{dotR1}) once more, we arrive to
\beq
\dot H= \frac{A}{12BR}~.
\eeq
From this the expression 
\beq
R\equiv 6\dot H + 12H^2 = \frac{5A}{6BR}-\frac{2r_1}{3}
\eeq
follows that cannot be satisfied for $\dot R\not= 0$.
Thus, the only remaining possibility is $B=0$ and then $A=0$ if $R\not= 0$.

So, all 
solutions of Eq. (\ref{trace}) under the condition (\ref{trace1}) are given either by 
$R\equiv 0$, or by $A=B=0$.
However, since $R\equiv 0$ is a particular solution of Eq. (\ref{trace1}) (or (\ref{wasans})), too, we notice that this case having zero measure in the space of initial
conditions for Eq. (\ref{tEOMtrace}) does not require special consideration further.

We finally note that one can extend the above proof to any metric which is space-homogeneous in the synchronous frame.

\section{Slow-roll approximation using (\ref{slowroll})}

\label{slowroll:app}

The slow-roll parameter $\epsilon$ defined in (\ref{slowroll}) and we note it down again here for references:
\begin{equation}
\epsilon=-\frac{\dot{H}}{H^{2}}=1-\frac{\mathcal{H}^{\prime}}{\mathcal{H}^{2}}\,.
\label{slowrollapp}
\end{equation}
During inflation which is a nearly dS expansion we have $\epsilon\ll1$ and use it as the small parameter for our approximations.
It is useful to compute
\begin{equation}
\epsilon^{\prime}=2\mathcal{H}\left(1-\epsilon\right)^{2}-\frac{\mathcal{H}^{\prime\prime}}{\mathcal{H}^{2}}\,,\label{epsilon}
\end{equation}
We require $\epsilon'\ll\Hc\epsilon$ which is the condition to have inflation last long enough.
The background scalar curvature and its conformal time derivatives become
\begin{equation}
	\begin{split}
	\bar R &=\frac{6\Hc^2}{a^2}(2-\epsilon)\,,~\bar R'=-2\Hc\epsilon\bar R-\bar R\frac{\epsilon'}{2-\epsilon}\approx-2\Hc\epsilon\bar R\,,\\
	\bar R'' &\approx -2\Hc\bar R (\epsilon'+\Hc\epsilon)+6\Hc^2\epsilon^2\bar R\approx-2\Hc^2\epsilon\bar R \approx \Hc\bar R'\,.
\end{split}
	\label{slowrollR}
\end{equation}
Substituting these expressions into (\ref{wasans}) one finds an important relation among our parameters
\begin{equation}
	r_1\approx\frac{6\Hc^2\epsilon}{a^2}\approx\frac{\bar R\epsilon}{2}\, .
	\label{r1eps}
\end{equation}
Further examining equation (\ref{00radnonlocal}) one finds
\begin{equation}
	\frac{M_P^2}{3M^2}\left[\frac32 R\dot H-3H\dot R-9M^2H^2\right]+\rho_r=O(\epsilon^2)+\rho_r=0\,.
	\label{00radnonlocaleps}
\end{equation}
As such, radiation contributes at least at the order $\epsilon^2$.

\section{Second order variation of action (\ref{actionour4inf}) around (\ref{wasans}) and (\ref{conditions})}

\label{s2:app}

We will actively use
\begin{equation}
	\Zc_1(\Box)=\frac{\Fc_R(\Box)-\Fc_1}{\Box-r_1}\,,\quad \Zc_2(\Box)=\frac{\Fc_R(\Box)-\Fc_1}{(\Box-r_1)^2}\,,\label{z1z2app}
\end{equation}
where both quantities are analytic in $\Box$ thanks to condition $\Fc_R'(r_1)=0$ (see (\ref{conditions})).
We also remind few sums:
\begin{equation*}
	\sum_{l=0}^{n-1}1=n\,,\quad\sum_{l=0}^{n-1}x^l=\frac{x^n-1}{x-1}\,,\quad\sum_{l=0}^{n-2}(l+1)x^l=-\frac{x^n-1}{(x-1)^2}+n\frac{x^{n-1}}{x-1}\,.
\end{equation*}
This allows to compute
\begin{eqnarray*}
	(\delta \Fc_R(\Box))\bar R&=& \Zc_1(\Box)(\delta\Box)\bar R\, ,\\
	\intdg{4}\bar R(\delta^2\Fc_R(\Box)|_{\delta^2\Box})\bar R&=& \intdg{4}\sum_{n=1}^\infty f_{Rn}\sum_{l=0}^{n-1}\bar R\Box^l(\delta^2\Box)\Box^{n-l-1}\bar R\sim\Fc_R'(r_1)=0\, ,\\
\intdg{4}\bar R(\delta^2\Fc_R(\Box)|_{\delta\Box\delta\Box})\bar R&=&\\
=\intdg{4}\sum_{n=2}^\infty f_{Rn}\sum_{\alpha+\beta\leq n-2}&\bar R&\Box^\alpha(\delta\Box)\Box^{n-\alpha-\beta-2}(\delta\Box)\Box^\beta\bar R=\\
&=& \intdg{4}\bar R(\delta\Box)\Zc_2(\Box)(\delta\Box)\bar R\, ,\\
\intdg{4}\bar R(\delta\Fc(\Box))\delta R&=& \intdg{4}\bar R (\delta \Box)\Zc_1(\Box)\delta R\, .
\end{eqnarray*}

The most tedious piece in varying (\ref{actionour4inf}) is the one with $\Fc_R$. Its variation contains 10 terms which explicitly can be written as
\begin{equation}
	\begin{split}
		\delta^2S_{\Fc_R}=\frac\lambda 2\intdg{4}&\left[\bar R\Fc_1\bar R\delta_g+(\delta^2R)\Fc_1\bar R+\bar R\Fc_1\delta^2 R+\bar R(\delta\Box)\Zc_2(\Box)(\delta\Box)\bar R\right.\\
			&+\frac h2(\delta R)\Fc_1\bar  R+\frac h2\bar R\Fc_R(\Box)\delta R+\bar R(\delta\Box)\Zc_1(\Box)\delta R\\
			&\left.+\frac h2\bar R\Zc_1(\Box)(\delta\Box)\bar R+(\delta R)\Zc_1(\Box)(\delta \Box)\bar R+(\delta R)\Fc_R(\Box)\delta R\right]\, ,
	\end{split}
	\label{d2SFR}
\end{equation}
where we use
\begin{equation*}
	\sqrt{-g}=\sqrt{-\bar g}\left(1+\frac h2+\delta_g+O(h_{\mu\nu}^3)\right),\quad\delta_g=\frac{h^2}8-\frac{h_{\mu\nu}^2}4
\end{equation*}
as well as the above derived relations.

To advance further we compute some quantities involving $h_{\mu\nu}$
\begin{equation}
	\begin{split}
	\delta g^{\mu\nu}&=-h^{\mu\nu}\, ,\quad
	\delta \Gamma^\rho_{\mu\nu}=\gamma^\rho_{\mu\nu}=\frac12(\nabla_\mu h^\rho_\nu+\nabla_\nu h^\rho_\mu-\nabla^\rho h_{\mu\nu})
\end{split}
	\label{gammarhomunu}
\end{equation}
and for the d'Alembertian acting on scalars
\begin{equation}
	\delta \Box=\delta(\nabla^\mu\pd_\mu)=-h^{\mu\nu}\nabla_\mu\pd_\nu-(\nabla^\mu h^\rho_\mu)\pd_\rho+\frac12(\pd^\rho h)\pd_\rho\, .
	\label{deltabox}
\end{equation}
We do not use the subscript ``$s$'' because no variations of other (acting on tensors for example) d'Alembertians are used in the paper.
Using this latter expression we can show utilizing the integration by parts
\begin{equation}
	\intdg{4}\bar R(\delta\Box)X=\intdg{4}\left[(\delta\Box \bar R)X-\frac12\bar R h (\Box-r_1)X\right]\, .
	\label{deltaboxX}
\end{equation}
This in combination with the definition (\ref{zetamain}) simplifies (\ref{d2SFR}) to
\begin{equation}
	\delta^2S_{\Fc_R}=\frac\lambda 2\intdg{4}\left[(2\delta_0-\bar R\delta_g)\Fc_1\bar R+\Fc_1(\delta R)^2+\zeta \Zc_2(\Box)\zeta\right]\, ,
	\label{d2SFRzeta}
\end{equation}
where
\begin{equation*}
	\delta_0=\bar R\delta_g+\frac h2\delta R+\delta^2R
\end{equation*}
is in fact the second order variation of the Einstein-Hilbert action.

Accounting the Einstein-Hilbert and Weyl tensor terms in (\ref{actionour4inf}) and using conditions (\ref{conditions}) we arrive to (\ref{delta2Sinf}).

\section{Linearized EOM and proof that $\zeta=0$ in slow-roll approximation}
\label{app:linear}
The variation of the trace equation (\ref{tEOMtrace}) reads 
\begin{equation}
\delta E=\left\{\left[\partial^{\mu}\bar{R}\partial_{\mu}+2r_{1}\bar{R}\right]\mathcal{Z}_{2}\left(\bar{\Box}_{k}\right)+
3\mathcal{\mathcal{F}}_{R}\left(\bar\Box_k\right)+\left(\bar{R}+3r_{1}\right)\mathcal{Z}_{1}\left(\bar\Box_k\right)\right\}\zeta=0\, ,
\label{dE}
\end{equation}
where $\zeta$ is defined in (\ref{zetamain}) and $\Zc_1,~\Zc_2$ are defined in Appendix~\ref{s2:app}.

The variation of the $({}^i_j)$-equation with $i\neq j$ in the system (\ref{tEOM}) yields
\begin{equation}
	\delta E^i_j=-2\lambda\frac{k^ik_j}{a^2}\left[\Fc_1(\bar R+3r_1)(\Phi-\Psi)+\Upsilon\right]+2\lambda c^i_j=0 \, . 
\label{eq342}
\end{equation}
The variation of the $({}^0_i)$-equation in the system (\ref{tEOM}) yields
\begin{equation}
	\!\!\delta E^0_i=2\lambda\frac{ik_i}{a^2}\left[2\Fc_1(\bar R+3r_1)(\Psi'+\Hc\Phi)-(\Upsilon'-\Hc\Upsilon)+\Fc_1\bar R'\Phi-\frac12\bar R'\Xi\right]+2\lambda c^0_i=0 \, . 
	\label{eq343}
\end{equation}
The variation of the $({}^0_0)$-equation in (\ref{tEOM}) yields
\begin{eqnarray}
\delta E^0_0&=&  \frac{2\lambda}{a^2}\bigg[-2\Fc_1(\bar R+3r_1)(3\Hc\Psi'+3\Hc^2\Phi+k^2\Psi)+3\Hc\Upsilon'-3\Hc'\Upsilon+k^2\Upsilon \nonumber \\
	& -& 3\Fc_1\bar R'(\Psi'+2\Hc\Phi)-\frac{\bar R'}2\Xi'+\frac12\left({\bar R''}+2\Hc \bar R'\right)\Xi\bigg]+2\lambda c^0_0=0 \, . 
	\label{eq344}
\end{eqnarray}
Here we have used the notations
\begin{eqnarray} 
\Upsilon=\frac{\Fc(\bar \Box_k)-\Fc_1}{\bar\Box_k-r_1}\zeta+\Fc_1\delta R_{\rm GI} 
\: \text{ and } \: 
\Xi=\frac{\Fc(\bar\Box_k)-\Fc_1}{(\bar \Box_k-r_1)^2}\zeta \, ,\nonumber
\end{eqnarray}
where $\delta R_{\rm GI}$ is the gauge invariant scalar curvature variation defined in (\ref{deltaRGI}).
Moreover
$c^\mu_\nu$ is found to be, 
\begin{eqnarray}
	c^i_j&=& \frac{k^ik_j}{a^2}\left(\Theta'+2\Hc\Theta+\frac{k^2}3\Omega\right)\, ,\qquad i\neq j \, , \nonumber\\
	c^0_i &=&  -\frac23\frac{ik_ik^2}{a^2}\Theta \, , \quad	c^0_0= \frac23\frac{k^4}{a^2}\Omega \, , \nonumber\\
	\Theta&=& \Omega'+2\Hc\Omega\, ,\quad\Omega=\Fc_{\rm C}(\bar \Box_k+6H^2)\frac{\Phi+\Psi}{a^2}\, .\nonumber
\end{eqnarray}
Here $\bar\Box_k$ is given in (\ref{frwboxsk}). {Note that the above expression for $c^i_j$ with $i\neq j$ is equivalent to one given in \cite{Koshelev:2016xqb} upon identical cancellations.}

To see explicitly that $\zeta=0$ is the only solution to (\ref{dE}) in the first order slow-roll approximation
let's assume for the beginning that
\begin{equation}
	\zeta=\sum_{i}\zeta_{\nu_{i}}\, ,\qquad
\bar{\Box}\zeta_{\nu_{i}}=w_{i}\zeta_{\nu_{i}}\,.\label{zeta0}
\end{equation}
Here $\bar{\Box}$ is the d'Alembertian operator for the exact dS
space and $w_i$ are constant. An assumption that an arbitrary function can be represented as a linear superposition of eigenfunctions of the d'Alembertian is exactly the condition \textit{(ii)} in the proposition proven in Section~\ref{sec:proposition} and justifications of the validity of this assumption are already presented there. Note that the Fourier decomposition of $\zeta$ can be taken
with respect to any choice of the d'Alembertian operator. The main property we utilize is the orthogonality of the Fourier eigenmodes.

In the dS limit $a\approx-\frac{1}{H\tau}$ and $\mathcal{H}\approx-\frac{1}{\tau}$ and we yield
for each $\zeta_{\nu_{i}}$ 
\begin{equation}
\zeta_{\nu_{i}}=\left(k\tau\right)^{3/2}\left[j_{ik}J_{\nu_{i}}\left(k\tau\right)+y_{ik}Y_{\nu_{i}}\left(k\tau\right)\right]\,,\label{zeta-isol}
\end{equation}
where $J_{\nu}\left(k\tau\right)$, $Y_{\nu}\left(k\tau\right)$
are Bessel and Neumann functions, $j_{ik},~y_{ik}$ are constants and
\begin{equation*}
\nu_{i}=\sqrt{\frac{9}{4}-\frac{w_{i}}{H^{2}}}\,.
\end{equation*}
Substituting explicit solution for $\zeta$ in (\ref{dE}) we get
in the leading slow-roll approximation
\begin{equation}
\epsilon\bar R^2\sum_{i}\mathcal{Z}_{2}\left(w_{i}\right)\left[-\frac \tau 6\zeta_{\nu_{i}}^{\prime}+\zeta_{\nu_{i}}\right]=-\sum_{i}\mathcal{W}\left(w_{i}\right)\zeta_{\nu_{i}}\,.\label{dEi}
\end{equation}
Here operator $\Wc$ is given in (\ref{Wbox}) and as explained below that definition must be an operator without eigenvalues. That is $\Wc(z)$ is non-zero for any complex $z$.
Using the recursion properties valid for Bessel functions (here $f_\nu(x)$ is either $J_\nu(x)$ or $Y_\nu(x)$ function)
\begin{equation}
\begin{aligned}
2\nu f_{\nu}\left(z\right)= & z\left[f_{\nu-1}(z)+f_{\nu+1}(z)\right]\,,\\
\frac{d}{dz}\left(f_{\nu}(z)\right)= & \frac{1}{2}\left(f_{\nu-1}(z)-f_{\nu+1}(z)\right)=\frac{\nu}{z}f_{\nu}\left(z\right)-f_{\nu+1}\left(z\right)=-\frac{\nu}{z}f_{\nu}\left(z\right)+f_{\nu-1}\left(z\right)\,,
\end{aligned}
\label{bess-prop}
\end{equation}
one can get
\begin{equation}
\begin{aligned}\zeta_{\nu_{i}}^{\prime}= & \frac k2\left(k\tau\right)^{3/2}\left[j_{ik}(J_{\nu_{i}-1}\left(k\tau\right)-J_{\nu_{i}+1}\left(k\tau\right))+y_{ik}(Y_{\nu_{i}-1}\left(k\tau\right)-Y_{\nu_{i}+1}\left(k\tau\right))\right]
	+\frac{3}{2\tau}\zeta_{\nu_{i}}
	\,.\end{aligned}
\label{zetap}
\end{equation}
We see from the last equation that the single derivative term with $\zeta_{\nu_i}$ in the trace equation can be recast in other $\zeta$-s but also in the same $\zeta$ with an extra time-dependent factor. The orthogonality of Bessel and Neumann functions together with the appearance of this extra $\tau$-factor proves $\zeta=0$ is the only solution to (\ref{dE}) in the leading slow-roll approximation.

Going further one can work out the same technique using the quasi dS scale factor and Hubble
parameter in terms of the conformal time 
\begin{equation}
a\sim-\frac{1}{H\tau}+\frac{\epsilon\left[\ln(-\tau)-1\right]}{H\tau}+{O}\left(\epsilon^{2}\right)\,,\quad\mathcal{H}=-\frac{1}{\tau}-\frac{\epsilon}{\tau}+{O}\left(\epsilon^{2}\right)\,.\label{aHeps}
\end{equation}
One can show upon construction of the $\epsilon$-corrected d'Alembertian and corresponding eigenmodes that that an appearance of new linearly independent functions upon computing the derivatives is the blocking issue for any solution apart from $\zeta=0$.


\bibliographystyle{JHEP}
\bibliography{nli}

\end{document}